\documentclass[pdflatex,sn-mathphys]{sn-jnl}

\jyear{2021}%

\theoremstyle{thmstyleone}%
\theoremstyle{thmstyletwo}%

\theoremstyle{thmstylethree}%
\usepackage{lineno}
\begin{document}

\title[VCG: Constraints from Supernovae
and Gravitational Wave Merger Events]{Variable Chaplygin Gas: Constraints from Supernovae, GRB and Gravitational Wave Merger Events}


\author[1]{\fnm{Ashley} \sur{Chraya}}\email{ashleychraya@gmail.com }

\author[2]{\fnm{Yuvraj} \sur{Muralichandran}}\email{yuvrajmuralichandran@gmail.com }

\author*[3]{\fnm{Geetanjali} \sur{Sethi}}\email{getsethi@ststephens.edu}

\affil[1]{\orgdiv{Department of Physics}, \orgname{Indian Institute of Science Education and Research}, \orgaddress{\city{Mohali}, \country{India}}}

\affil[2]{\orgdiv{Department of Physics and Astronomy}, \orgname{University of Potsdam}, \orgaddress{\city{Potsdam}, \postcode{10587}, \country{Germany}}}

\affil[3]{\orgdiv{Associate Professor,Department of Physics}, \orgname{St. Stephen's College, University of Delhi}, \orgaddress{\street{Street}, \city{Delhi}, \postcode{110007}, \state{Delhi}, \country{India}}}


\abstract{We investigate the cosmological constraints on the Variable Chaplygin gas model from the latest observational data: SCP Union 2.1 compilation dataset of Type Ia supernovae (SNe Ia), Pantheon sample of SNe Ia, Platinum Sample of Gamma Ray Bursts (GRB) and GWTC-3 of gravitational wave merger events. Variable Chaplygin gas is a model of interacting dark matter and dark energy, which interpolates from a dust-dominated era to a quintessence-dominated era.
The Variable Chaplygin gas model is shown to be compatible with Type Ia Supernovae and gravitational merger data. We have obtained tighter constraints on cosmological parameters $B_s$ and $n$, using the Pantheon sample. By using the Markov chain
Monte Carlo (MCMC) method on the Pantheon sample, we obtain $B_s$=0.108 $\pm$ 0.034, n=1.157 $\pm$ 0.513 and $H_0$=70.020 $\pm$ 0.407, for GRBs, we obtain $B_s$=0.20 $\pm$ 0.11, n=1.45 $\pm$ 1.40 and $H_0$=70.41 $\pm$ 0.67} and on GWTC-3, we obtain $B_s$=0.130 $\pm$ 0.076, n=0.897 $\pm$ 1.182 and $H_0$=69.838 $\pm$ 3.007. The combined constraints from the above data sets are $B_s$=0.11 $\pm$ 0.03, n=1.14 $\pm$ 0.36 and $H_0$=70.34 $\pm$ 0.61




\maketitle

\section{Introduction}\label{sec1}

Most of the time, the observations have the power to push the well-established field to its most stringent tests. This is especially true in modern cosmology, and one such observation was of the supernovae type-Ia (SN-Ia) by two independent groups in 1998 \citep{Perlmutter_1999,riess1998observational}, which established that the expansion of the universe is accelerating and two-thirds of total energy density is of dark energy component with negative pressure. This was followed by massive experimental efforts to find other independent observational means to verify the expansion of the universe theory. Through a series of independent findings such as the Baryon Acoustic Oscillations (BAO), Cosmic Microwave Background Radiation (CMBR) \citep{Spergel_2003, miller1999measurement}, large-scale structure (LSS) \citep{bahcall1999cosmic}, and Gamma-ray Bursts (GRBs), the accelerated expansion of the universe and the existence of dark energy gained traction in cosmology.

Regardless of the limitations of cosmological models based on the cosmological principle, numerous cosmological models have been extremely successful in explaining the observed universe. The simplest of such models, the Friedmann-Lemaître-Robertson-Walker (FLRW) model, assumes cosmological principle, Einstein’s general relativity and the Universe is composed of Baryonic matter and radiation. Assuming the universe is homogeneous and isotropic, FLRW metric is,
\begin{equation}
d s^{2}=d t^{2}-a^{2}(t)\left[\frac{d r^{2}}{1-k r^{2}}+r^{2}\left(d \theta^{2}+\sin ^{2} \theta d \phi^{2}\right)\right]
\end{equation}
where k= curvature of the space which can take values $k=0, \pm 1$ and a(t) is the scale factor. Even though this model was extremely successful there are many aspects of our observed universe which cannot be captured by such a simple model. For instance, it is not possible to explain the accelerated expansion if assumptions of FLRW model i.e. cosmological principle, Einstein's general relativity and only Baryonic matter and radiation is taken into consideration. Subsequently, other models were explored which can be broadly classified into two classes. One class of models involve reforming the geometry part of the Einstein’s equations. This includes generalisation of the gravitational action such as f(R)-gravity and higher dimension spacetime. The other class of models alter the matter component of the Universe in the Einstein’s equations. If it is assumed that the universe is composed only of Baryonic matter and radiation as it was assumed in FLRW model, due to the gravitational pull of these components expansion rate of the universe would have slowed down, therefore, the need for an additional entity to explain the accelerated expansion of the Universe was realised. In this approach, exotic matter with equation of state such that they don't follow strong energy conditions is added to the mass distribution of the Universe. This conclusion is easily followed by second Friedmann equations. One possible approach to constructing a viable model for dark energy is to associate it with a slowly evolving and spatially homogenous scalar field $\phi$, called “quintessence” \citep{PhysRevD.37.3406, PhysRevLett.82.896,guo2007parametrizations} or two coupled fields \citep{bento2002two}. Over a large class of potentials, the Quintessence model gives the energy density convergent to its present value for a wide range of initial conditions in the past and exhibit tracker behaviour \citep{sahni2000case,padmanabhan2003cosmological}. Some of the models based on adding exotic matter are quintessence, k-essence, tachyons, barotropic fluid etc. Quintessence, k-essence are scalar field models, whereas Dark Energy models include barotropic fluids whose
pressure is a function of energy density, $P = f (\rho)$. The relationship between pressure and energy density determines the dynamics of the fluid.

Observational results of SN Ia, as well as the anisotropy of the CMBR power spectrum and clustering estimates, show that our universe is also composed of dark matter, in addition to dark energy, Baryonic matter and radiation. Among the various models, the one that is most favourable is the standard model of cosmology or $\Lambda$-Cold Dark Matter ($\Lambda$CDM) model, where $\Lambda$ represents the cosmological constant accounting for the vacuum energy or the energy density of space. However, this model faces some serious problematic issues such as the ﬁne tuning (unconventional small value) \citep{RevModPhys.75.559,RevModPhys.61.1} and cosmic coincidence problems (why the dark matter and the dark energy are of the same order today although the universe is in a speedy expansion phase?). Similarly the "Quintessence” models also suffer from fine tuning problem. 

The cosmological constant model has been transformed into a dynamical form in several ways to remove these problematic issues. We can describe an interaction between dark matter and dark energy in order to execute the dynamical form of the cosmological constant dark energy \citep{barrow2006cosmologies,gonzalez2006dynamics,boehmer2008dynamics,jamil2009constraining}. Hence, as an alternative to both the cosmological constant and quintessence, the accelerating expansion of the universe may be explained by introducing a cosmic fluid component with an exotic equation of state known as Chaplygin gas (CG) \citep{kamenshchik2001alternative} which is an example of barotropic fluid. Moreover, dark matter and dark energy has no laboratory evidence for its existence directly which forces us to find a model in which these two dark components are different manifestations of a single cosmic fluid. CG model unifies the CDM and the $\Lambda$ models' features into a single component with an exotic equation of state.  This versatility of the models built on the Chaplygin gas becomes an attractive feature as these models can explain both dark energy and dark matter with a single component, which makes it unified dark matter/energy (UDME) model. The CG model has been further extended also to the generalized \citep{bilic2002unification,saadat2014vol}, modiﬁed \citep{debnath2004role}, Variable\citep{chimento1996exact} and the extended \citep{kahya2015higher} forms. The CG model arises from the string Nambu–Goto action
in the light-cone coordinate \citep{jackiw2000particle,pedram2008quantum}. For the generalized Chaplygin gas (GCG) model, the
action can be written as a generalized Born–Infeld form \citep{bento2002generalized}. These models have been found to be consistent with the SNe Ia data \citep{gong2004constraints}, CMB peak locations
\citep{carturan2003cosmological} and other observational tests like gravitational lensing, cosmic age of old high redshift objects etc. \citep{dev2004constraints},
as also with some combination of some of them \citep{PhysRevD.69.083503}. This model has been shown to fit within the standard structure formation scenarios \citep{bilic2008transient,bento2002generalized,fabris2002density}. Generalized Chaplygin gas model is shown to be consistent with the constraints provided by 21 cm absorption line from EDGES detection along with CMB probe \citep{Yang_2019}. Therefore, the Generalized Chaplygin gas model seems
to be a good alternative to explain the accelerated expansion of the universe. As the original Chaplygin gas model produced oscillations or exponential blowup of matter power spectrum that are inconsistent with observations, various modified models have been considered. In this paper we consider the variable Chaplygin gas model that  was proposed \cite{VCGPrime} and constrained using SNeIa “gold” data \cite{Sethi_2006}. There are also hybrid models like Viscous Generalized Chaplygin Gas (VGCG), that were proposed to tackle the late accelerated expansion of the Universe.  This kind of hybrid models was originally proposed by \cite{Zhai:2005mu} and are able to avoid causality problems that arise when only dissipative fluids are considered.

The parameters of {\it{variable}} Chaplygin gas model (VCG) has been constrained using statistical analysis of SNeIa (Supernova Type Ia) Union 2.1 dataset from Supernova Cosmology Project, Pantheon+ Dataset from SH0ES collaboration and with the 03 run of GWTC. Formal introduction to the {\it{variable}} Chaplygin gas model is provided briefly in section \ref{sec2} and with description of the Pantheon and GWTC-3 datasets used for constraining the parameters, in section \ref{sec3} and the description of GRBs in section \ref{3D correlation}. The statistical formalism to constrain parameters is given in section \ref{sec4}. The analysis on the performance of VCG model with dataset and the constraints obtained on the parameters of the model is discussed in section \ref{sec5}, followed by conclusion of the study in section \ref{sec6}

\section{VariableChaplygin Gas Model}\label{sec2}
The equation of state for the Chaplygin gas is $P =$-$A/\rho$ where A is a positive constant. A generalized Chaplygin gas model is characterised by an equation of state

\begin{equation}\label{chaplygineq}
P_{ch}=-\bigg(\frac{A}{\rho_{ch}^{\alpha}}\bigg)
\end{equation}

where $\alpha$ is a constant such that $0<\alpha\leq1$. 
The Chaplygin gas model corresponds to $\alpha =1$. Taking time component of the energy-momentum  conservation equation $T_{, \mu}^{\mu \nu}=0$, we obtain the continuity equation,
\begin{equation}
\frac{\partial \rho}{\partial t}+3 H(p+\rho)=0
\label{continuityeq}
\end{equation}
where H is the Hubble parameter, $H=\frac{\dot{a}}{a}$,  $\rho$ = total density of Universe. Using equation (\ref{continuityeq}), the energy density of the Chaplygin gas evolves as \cite{PhysRevD.66.043507}

\begin{equation}
\rho{}_{ch} = \bigg( A+\frac{B}{a^{3(1+\alpha)}} \bigg) ^ {\frac{1}{1+\alpha}}
\label{chaplyginrhoarelation}
\end{equation}

where $a$ is the scale factor in the current epoch of the universe, and $B$ is the constant of integration. In the earlier epochs i.e. $a\ll1$,  equation (\ref{chaplyginrhoarelation}) translates to $\rho \propto a^{-3}$, therefore, the Chaplygin gas behaves like Cold Dark Matter (CDM) and for late epochs where the scale factor of the Universe is $a\gg1$ equation (\ref{chaplyginrhoarelation}) translates to $p = -\rho = constant$, therefore, the behavior of the Chaplygin gas is like the cosmological constant at alter times thus leading to an accelerated expansion of the Universe. Thus, Chaplygin gas equation of state leads to
a component which behaves as non relativistic matter at early times and as cosmological constant equal to $8 \pi G A^{1 /(1+\alpha)}$ at later stage. 

The models based on Chaplygin gas haven't faltered the interest it gained among researches as they have shown promise in the past. However, the Chaplygin gas
model produces oscillations or an exponential blowup of matter power spectrums that are inconsistent with
observations \citep{sandvik2004end}. This instability may be avoided by taking into account the combined effect of shear and rotation, which slows down the collapse with respect to the simple spherical collapse model \citep{del2013shear}. Subsequently a modification of Chaplygin gas as proposed is VariableChaplygin gas (VCG) \citep{VCGPrime}. The VCG is a modification of the Chaplygin gas model where the equation of the state of the Chaplygin Gas is free flowing across the epochs. 

The VCG emerges from the dynamics of a generalized d-brane in a (d + 1, 1)
spacetime and can be described by a complex scalar field $\phi$ whose action can be written as a generalized Born-Infeld action \citep{bento2002generalized}. Considering a Born-Infeld Lagrangian \citep{sen2002tachyon}
\begin{equation}
\mathcal{L}_{\mathrm{BI}}=V(\phi) \sqrt{1+g^{\mu \nu} \partial_{\mu} \phi \partial_{\nu} \phi}
\end{equation}
where $V(\phi)$ is the scalar potential. In a spatially flat FLRW universe, the energy density and pressure are given by $\rho=V(\phi)\left(1-\dot{\phi}^{2}\right)^{-1 / 2}$ and $P=-V(\phi)\left(1-\dot{\phi}^{2}\right)^{1 / 2}$, respectively. Thus, the corresponding equation of state of CG is given by\begin{equation}
P=-\frac{V^{2}(\phi)}{\rho}
\end{equation}
One can rewrite the self-interaction potential as a function of the cosmic scale factor: $V^2(\phi) =A(a)$. Thus, VCG is characterised by the equation of state:
\begin{equation}
P{}_{ch}=-\frac{A(a)}{\rho_{ch}}
\label{eq:equationofstate}
\end{equation}
where $A(a)=A_{0}a^{-n}$ is a positive function of the cosmological scale factor $a$, $A_{0}$ and $n$ are constants. Using the energy conservation equation, equation (\ref{continuityeq})  in a flat Friedmann-Robertson-Walker universe and equation (\ref{eq:equationofstate}), the VariableChaplygin gas density evolves as:
\begin{equation}
\label{vcgevolution}
\rho{}_{ch}=\sqrt{\frac{6}{6-n}\frac{A{}_{0}}{a{}^{n}}+\frac{B}{a{}^{6}}}
\end{equation}

where B is a constant of integration. Using Einstein field equations, $G_{\alpha\beta}=8\pi G T_{\alpha\beta}$, and FLRW metric, we obtain Friedmann equations
\begin{equation}
\begin{aligned}
H^{2} &=\frac{8 \pi G}{3} \rho-\frac{k}{a^{2}} \\
\frac{\ddot{a}}{a} &=-\frac{4 \pi G}{3}(\rho+3 p)
\end{aligned}
\end{equation} 
where $\rho$ is baryonic matter density. If dark energy component is taken into consideration, first Friedmann equation gives the expansion rate of the Universe in terms of matter and radiation density, $\rho$, curvature, $k$, and the cosmological constant, $\Lambda$, as
\begin{equation}
\label{freidmann}
H^{2} \equiv \bigg( \frac{\dot{a}}{a}\bigg)^{2} = \frac{8\pi G}{3}\rho-\frac{k}{a^{2}}+\frac{\Lambda}{3}
\end{equation}
Assuming, critical density $\rho_{c} = \frac{3{H^{2}}}{8{\pi}G}$ and density parameter $\Omega=\frac{\rho}{\rho_c}$, first Friedmann equation reads as 
\begin{equation}
    \Omega_{b}(a) + \Omega_{k}(a) =1
\end{equation}
This relation extends directly to other models with several components. Here, taking baryonic matter and radiation into account, Friedmann equation reads as,
    \begin{equation}
        H^2=H_0^{2} (\Omega_{m,0} a^{-3}+ \Omega_{\gamma,0} a^{-4}+\Omega_{k,0} a^{-2})
    \end{equation}
where $\Omega_{m,0}+ \Omega_{\gamma,0}+ \Omega_{k,0}=1$. Every other component can be added when its behaviour with scale factor is known. 

After reducing Eq. (\ref{freidmann}) to the case of the spatially flat Universe,
\begin{equation}
H{}^{2}=\frac{8\pi G}{3}\rho
\end{equation}
where $H\equiv \dot{a}/a$ is the Hubble parameter. Therefore the acceleration condition $\ddot{a} > 0$ is equivalent to
\begin{equation}
\left(\frac{12-3n}{6-n}\right)a^{6-n}>\frac{B}{A_{0}}
\end{equation}
In order to incorporate accelerated expansion of the Universe, the necessary condition is $n<4$.

For $n=0$, the original Chaplygin gas behaviour is restored. The gas initially behaves as dust-like matter ($\rho_{ch} \propto a^{-3}$) and later as a cosmological constant ($p = -\rho = constant$). However, in the present case of VCG the universe tends
to be a quintessence-dominated ($n > 0$) \citep{hannestad2002probing,guo2005parametrization} or phantom-dominated one ($n < 0$) \citep{caldwell2003phantom} with constant equation of state parameter $w = $-$1 + n/6$. The first term on the right hand side
of Eq. (\ref{vcgevolution}) is initially negligible so that the expression Eq. (\ref{vcgevolution}) can approximately be written as
$\rho \sim a^{-3}$ , which corresponds to a universe dominated by dust-like matter.

The present value of energy density of the Variable Chaplygin gas 
\begin{equation}
\rho_{cho}=\sqrt{\frac{6}{6-n}A_{0}+B}
\end{equation}
\\
where $a_{0}=1$. Defining a parameter, $B_s$,
\begin{equation}
B_s=\frac{B}{6A_{0}/(6-n)+B}
\end{equation}
\\
the energy density becomes 
\begin{equation}
\rho_{ch}(a)=\rho_{ch0}\left[\frac{B_s}{a^{6}}+\frac{1-B_s}{a^{n}}\right]^{1/2}
\label{eq:energydensity}
\end{equation}

\subsection{Model}
The Friedmann equation, using equation (\ref{eq:energydensity}) and $a=\frac{1}{1+z}$ for a Variable Chaplygin gas, becomes
\begin{equation}
\begin{split}
 H^{2}=&\frac{8\pi G}{3}\bigg\{\rho_{r0}(1+z)^{4}+\rho_{b0}(1+z)^{3}+\\&\rho_{ch0}\Big[B_s(1+z)^{6}+(1-B_s)(1+z)^{n}\Big]^{1/2}\bigg\}
\label{eq:friedmanvcg}
\end{split}
\end{equation}
where $\rho_{r0}$ and $\rho_{b0}$ are the present values of energy densities of radiation and baryons, respectively. Using\footnote{We have used the fact that for a flat Universe, $\Omega_{b0}+\Omega_{r0}+\Omega_{ch0}=1$, i.e the total matter density sums up to unity.}
\begin{equation}
\frac{\rho_{r0}}{\rho_{ch0}}=\frac{\Omega_{r0}}{\Omega_{ch0}}=\frac{\Omega_{r0}}{1-\Omega_{r0}-\Omega_{b0}}
\end{equation}
\\
and
\begin{equation}
\frac{\rho_{b0}}{\rho_{ch0}}=\frac{\Omega_{b0}}{\Omega_{ch0}}=\frac{\Omega_{b0}}{1-\Omega_{r0}-\Omega_{b0}},
\end{equation}
\\
Equation (\ref{eq:friedmanvcg}) becomes,
\begin{equation}
H^{2}=\Omega_{ch0}H_{0}^{2}a^{-4}X^{2}(a),
\label{FreidmannVCG}
\end{equation}
where
\begin{equation}
\begin{split}
X^{2}(a)=&\frac{\Omega_{r0}}{1-\Omega_{r0}-\Omega_{b0}}+\\&\frac{\Omega_{b0}a}{1-\Omega_{r0}-\Omega_{b0}}+a^{4}\bigg(\frac{B_s}{a^{6}}+\frac{1-B_s}{a^{n}}\bigg)^{1/2}
\end{split}
\end{equation}

\section{Datasets}\label{sec3}
SNe Ia are crucial to understand expansion of the universe. We examine the parameters of the VariableChaplygin gas model using a statistical analysis of the most recent SNe Ia data from the Pantheon Sample \citep{scolnic2018complete}, the Supernova Cosmology Project (SCP) Union 2.1 compilation \citep{suzuki2012hubble} and GWTC-3 gravitational waves dataset. Pantheon sample is composed of 1048 SNe Ia data whose redshifts span from z = 0.01 up to z = 2.26. SCP Union 2.1 compilation is composed of 580 SNe Ia data whose redshifts span from z=0.623 up to z=1.415 providing redshift, distance
moduli and associated errors in distance moduli. The gravitational merger events are obtained from the GWOSC (Gravitational Wave Open Science Center) which has the events obtained from detectors at LIGO Hanford, LIGO Livingston and LIGO Virgo. The events are collected across the three runs: O1 (from 12 September 2015 to 19 January 2016), O2 (from 30 November 2016 to 25 August 2017) and the O3 runs, O3a (from 1 April 2019 to 30 September 2019) and O3b (from 1 November 2019 to March 2020). The data set consist of 90 confirmed events in the GWTC-3 \citep{LIGODATA} \cite{2021arXiv211103606T}. The catalog contains events whose sources are black hole binary mergers up to a redshift of 0.90. The merger events from the GWOSC dataset (from O1 run to the recent O3 run) are used to constrain the parameters of the Variable Chaplygin gas model. The redshift data from these events were taken to predict luminosity distance and distance modulus using Variable Chaplygin gas model and compare with the luminosity distance obtained from the merger events using Bayesian inference.

\section{Statistical Methodologies}\label{sec4}
In this work, we analyse luminosity distance cosmological  observable. Luminosity distance show an explicit dependence on the cosmological model under consideration. Hence with this observable we can compare the model with the experimentally observed value of such observable.

\begin{figure}[htp]
 \begin{center}
     \includegraphics[height=9cm,width=9cm]{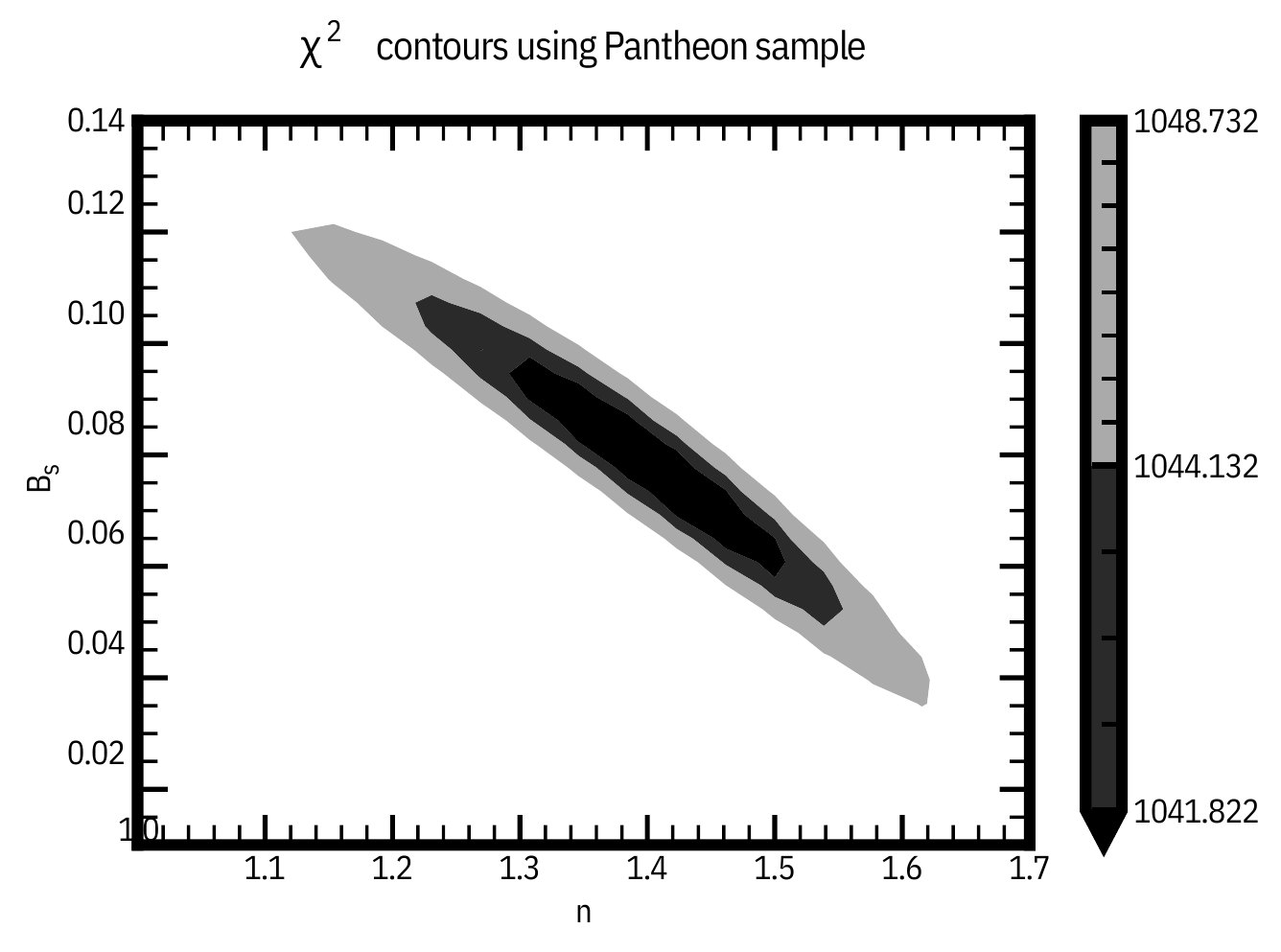}
    \caption{$\chi^2$ contour plot obtained for the Pantheon sample. Gradients represents levels of $\sigma$ confidence interval range obtained from the $\chi^2$ goodness test with two parameters $B_s$ and $n$ varying and fixed Hubble Parameter. Results are shown in table \ref{table2}.}
    \label{ContourPantheon}
\end{center}
\end{figure}

\begin{figure}[htp]
\begin{center}
    \includegraphics[height=9cm,width=9cm]{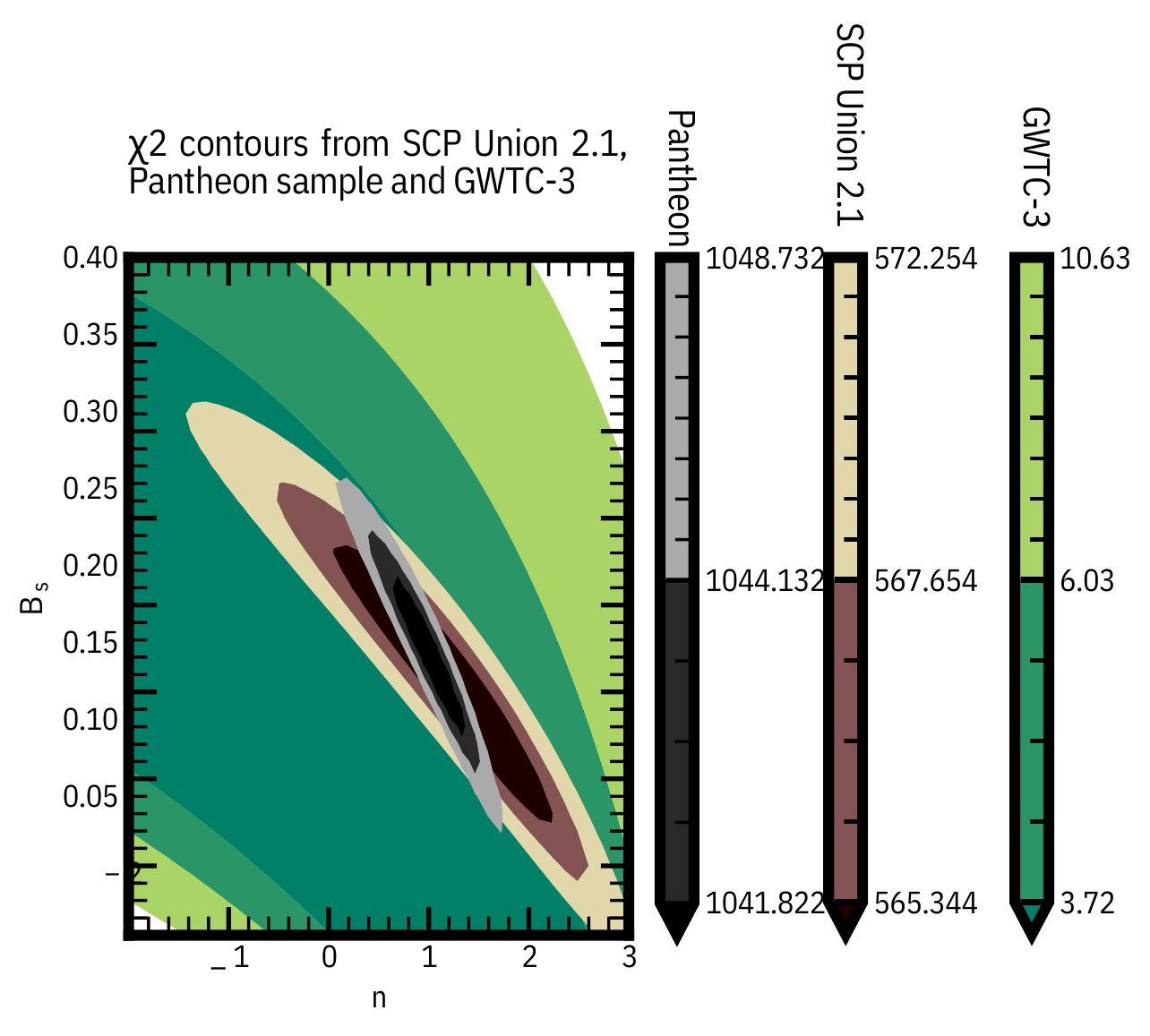}
     \caption{$\chi^2$ contour plots obtained for the Pantheon sample, SCP Union 2.1 compilation and GWTC-3 dataset with an overlap of confidence intervals for all three observational samples. Results are shown in table \ref{table2}.}
\label{chisqPantheonSCPGW} 
\end{center}
\end{figure}

We have used the SNe-Ia and gravitational waves to constrain the parameters of the Variable  Chaplygin gas model. Using the Friedmann equation, in a flat Universe, the luminosity distance is expressed as
\begin{equation}
d_{L}(z, \mathbf{p})=c(1+z)\int_{0}^{z} \frac{dz'}{H(z',\mathbf{p})}
\label{eq:ld_rel}
\end{equation}

where, $\{\mathbf{p}\}$ denotes the set of all parameters describing the cosmological model and $H(z, \mathbf{p})$ denotes the Hubble parameter as defined in the model chosen. In our case we have taken into consideration the contributions from radiation and baryons in addition to the Chaplygin gas. Luminosity distance can also be expressed using (\ref{FreidmannVCG}) and $a=\frac{1}{1+z}$ as

\begin{equation}\label{DLeq}
d_{L}=\frac{c(1+z)}{H_{0}} \int_{0}^{z} \frac{d z}{\Omega_{c h 0}^{1 / 2} X(z)}
\end{equation}
The theoretical predicted value of distance modulus is obtained by \begin{equation}
\mu_{p}^{(l)}=5 \log \left[\frac{d_{L}\left(z_{l}\right)}{\mathrm{Mpc}}\right]+25
\end{equation}
We can determine a best fit to the set of parameters by using a
$\chi^{2}$ goodness fit test,
\begin{equation}
\chi^{2}(\mathbf{p})=\sum_{l} \frac{\left[\mu_{p}^{(l)}\left(z_{l} \mid \mathbf{p}\right)-\mu_{0}^{(l)}\left(z_{l}\right)\right]^{2}}{\sigma_{l}^{2}}
\end{equation}
where $\mu_{p}^{(l)}(z_{l} \mid \mathbf{p}$) is the theoretical predicted value of distance modulus at redshift $z_{l}$, given the set of parameters p, and the sum is over all the observed data. Here $\sigma_{l}$ is the dispersion of the distance modulus due to intrinsic and observational uncertainties. In the case of SNe Ia, the errors $\sigma_{l}$  are Gaussian and hence, probability density function for
the parameters ($B_s$, $n$ and $H_0$) is
\begin{equation}
p(\mathbf{s}) \propto \exp \left(-\frac{\chi^{2}}{2}\right)
\end{equation}
with the parameter limits $0 \leq B_s \leq 1$ and $ n \leq 4$. The probability density function of a parameter $\mathbf{p}_i$ is obtained by integrating over all possible values of other parameters. To reduce the computation time, we can analytically integrate over Hubble constant and hence with modified $\chi^2$ statistics the fit is obtained by minimizing:

\begin{equation}\label{chisqmodified}
\chi^{2}=\sum_{i}\bigg[\frac{\mu_{th}^{i}-\mu_{obs}^{i}}{\sigma_{i}}\bigg]^{2}-\frac{C_{1}}{C_{2}}\bigg(C_{1}+\frac{2}{5}\ln10\bigg)-2\ln\mathit{h},
\end{equation}

here, 
\begin{equation}
C_{1}\equiv\sum_{i}\frac{\mu_{th}^{i}-\mu_{obs}^{i}}{\sigma_{i}^{2}},
\end{equation}
and
\begin{equation}
C_{2}\equiv\sum_{i}\frac{1}{\sigma_{i}^{2}}.
\end{equation}
where $\sigma_{i}$ is the error associated with observed distance modulus $\mu_{obs}$. The modified distance modulus in the above relation is

\begin{equation}
\mu_{th}=5\log\frac{H_{0}d_{L}}{ch}+42.38
\label{DMeq}
\end{equation}
$H_{0}$ is the value of Hubble parameter at present, $c$ is the speed of light. The probability distribution function of the estimated parameters (excluding h) is now
obtained by using the modified $\chi^2$. It is straightforward to check that the derivative of equation (\ref{chisqmodified}) with respect to h is
zero; hence our results are independent of the choice of h. We take $h = 0.70$. By minimizing the modified $\chi^2$ statistics we have found best fit for $B_s$ and $n$. We have obtained $\chi^2$ contour plot with confidence intervals for the Pantheon sample, see Fig. \ref{ContourPantheon}. As we are fitting for two parameters, we use  $\chi^2_{0.68} = \chi^2_{min} + 2.30$, $\chi^2_{0.95} = \chi^2_{min} + 4.61$, $\chi^2_{0.99} = \chi^2_{min} + 9.21$ for the value corresponding to 1 $\sigma$, 2 $\sigma$, 3 $\sigma$ respectively. We have also obtained $\chi^2$ contour plots with confidence intervals for the Pantheon sample, SCP Union 2.1 compilation and GWTC-3 together, see Fig. \ref{chisqPantheonSCPGW}. We are able to constrain the parameters of the model more tightly with the Pantheon sample than SCP Union 2.1 compilation. Therefore, for further analysis in this paper we will use the Pantheon sample only. For gravitational wave dataset, the errors in the observables are non-Gaussian, but we applied the above analysis for the gravitational wave dataset taking average of the upper and lower error bars. As the errors on the gravitational waves observables are too large we are not able to constrain the parameters of the model with it. Though, we are able to show that parameter values using $\chi^2$ by the Pantheon sample and GWTC-3 are consistent.

\begin{figure*}[ht!]
  \centering
    \includegraphics[height=5cm, width=11.5cm]{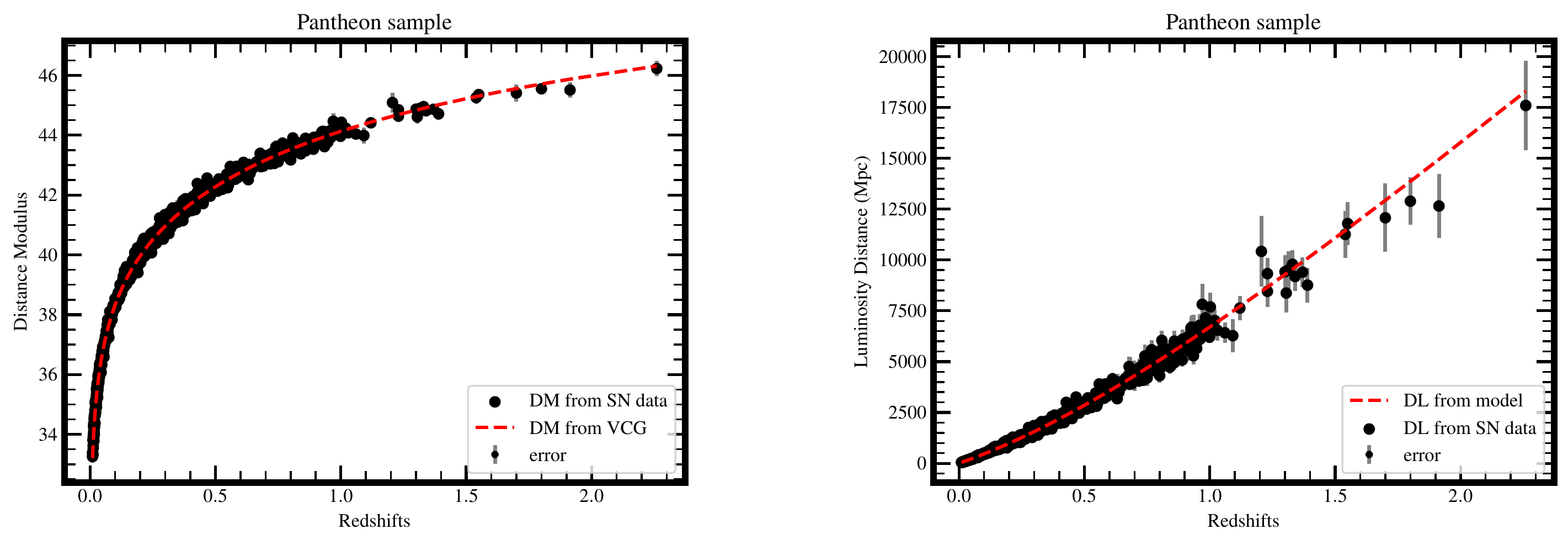}
    \includegraphics[height=5cm, width=11.5cm]{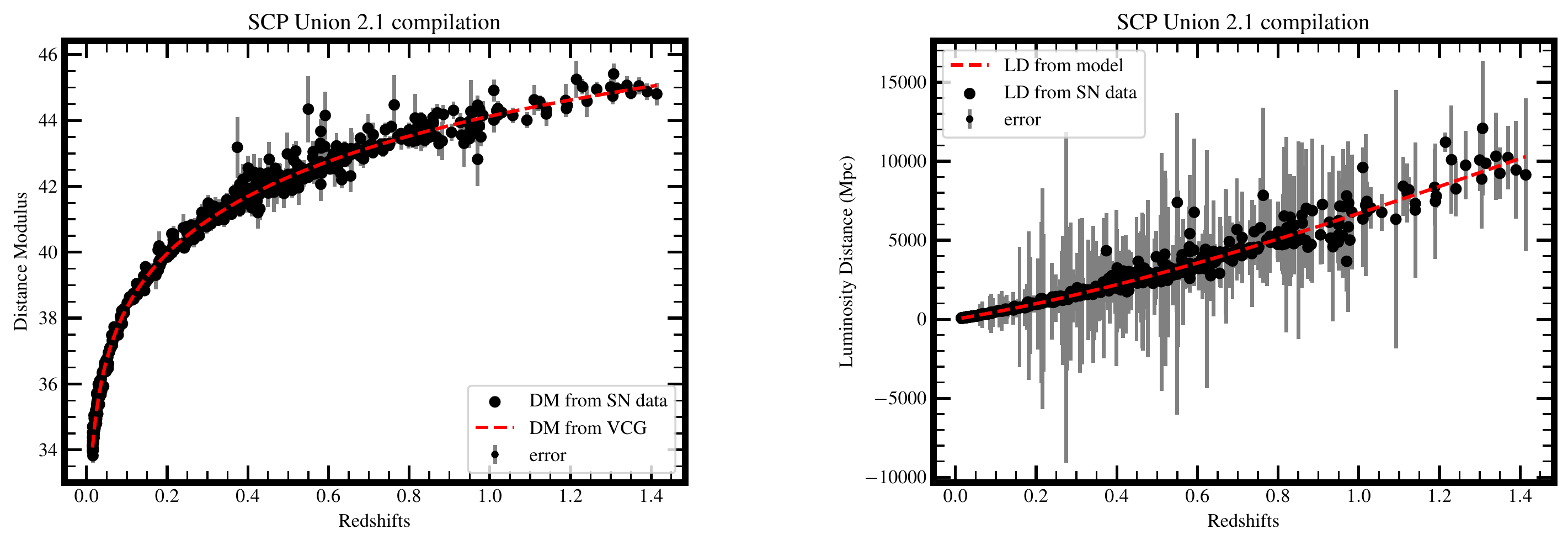}
     \includegraphics[height=5cm, width=11.5cm]{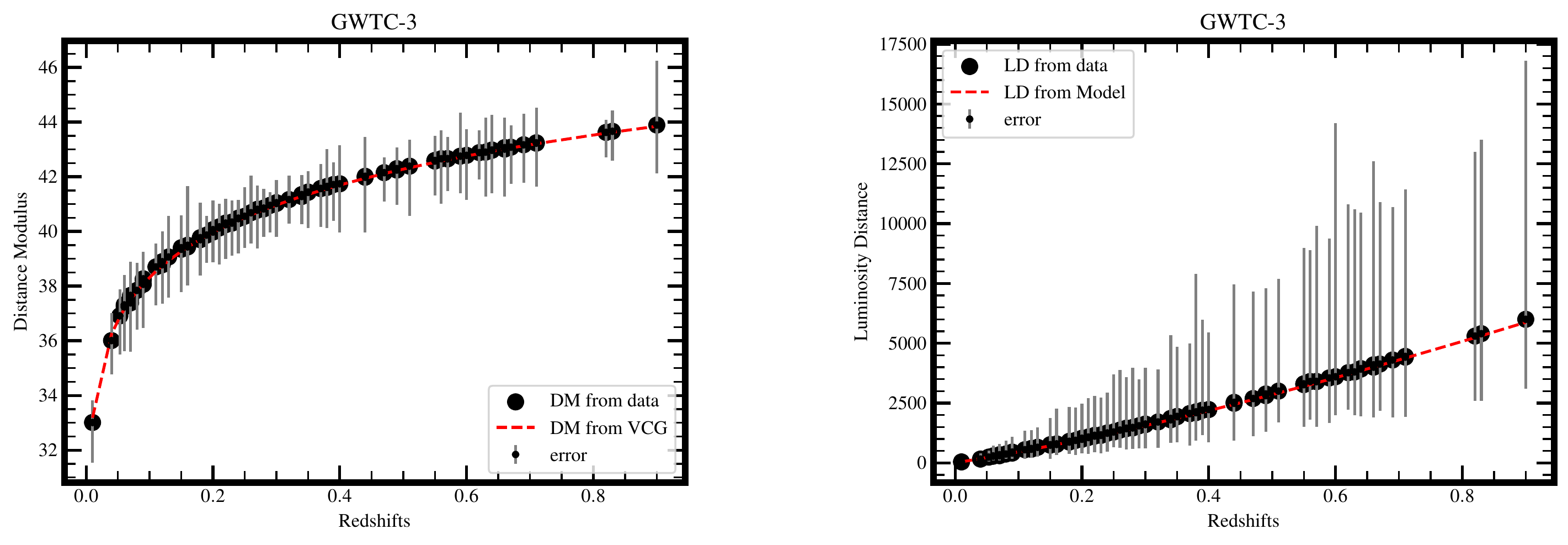}
  \caption{ Viability of Variable Chaplygin gas model with Pantheon, SCP Union 2.1 and GWTC-3 datasets. Variation of theoretical prediction of distance modulus (DM) (left) and luminosity distance (DL or LD) (right) from Variable Chaplygin gas model with redshift where the parameters ($B_s$, $n$) are determined by $\chi^{2}$ goodness test.  This shows model fits well with the Pantheon sample as well as with GWTC-3 dataset.}
  \label{PS Plot}
\end{figure*}

The theoretical luminosity distance and distance modulus from the equation (\ref{DLeq}) and equation (\ref{DMeq}) is compared to the observed values of distance luminosity and  distance modulus from the Pantheon, SCP Union 2.1 and GWTC-3 dataset respectively, see Fig. \ref{PS Plot}. We have shown that the model fits well and the viability of the model with all three samples. This shows Variable Chaplygin gas model (interacting dark energy and dark matter model) is capable of explaining the observed expansion of the universe.

\begin{figure}[htp]\label{H0estimate}
\begin{center}
    \includegraphics[height=6.0cm,width=7.0cm]{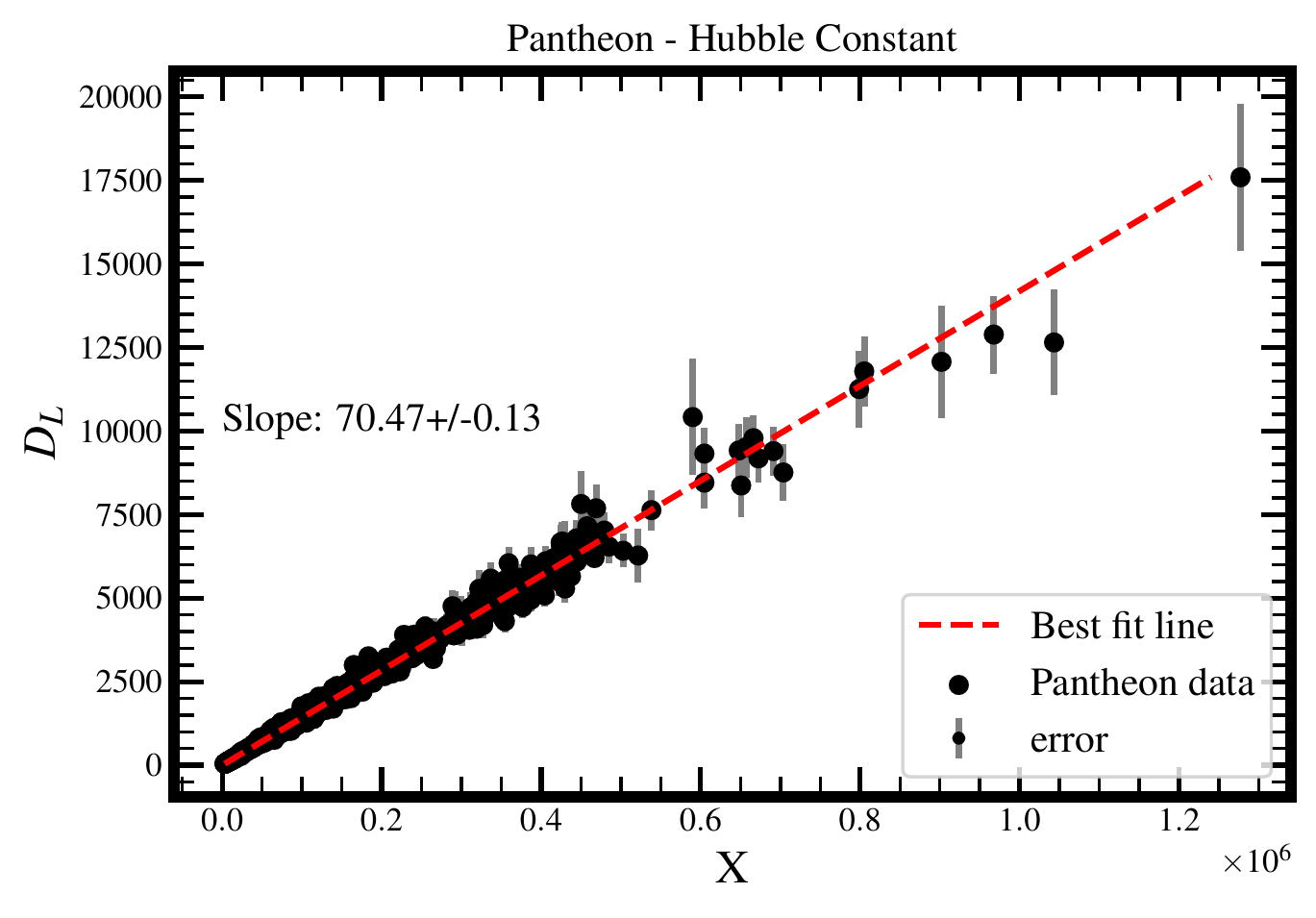}
    \includegraphics[height=6.0cm,width=7.0cm]{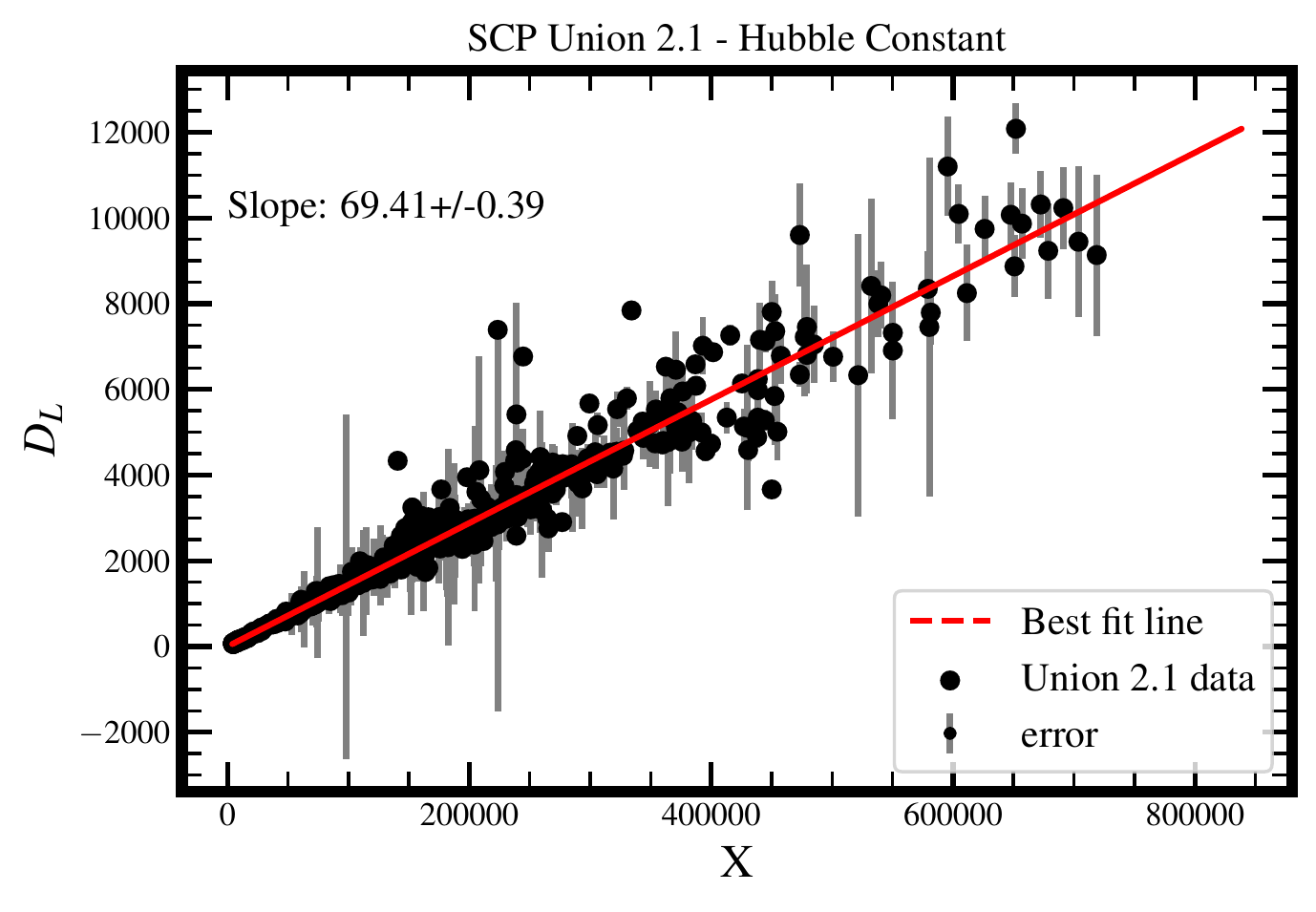}
    \includegraphics[height=6.0cm,width=7.0cm]{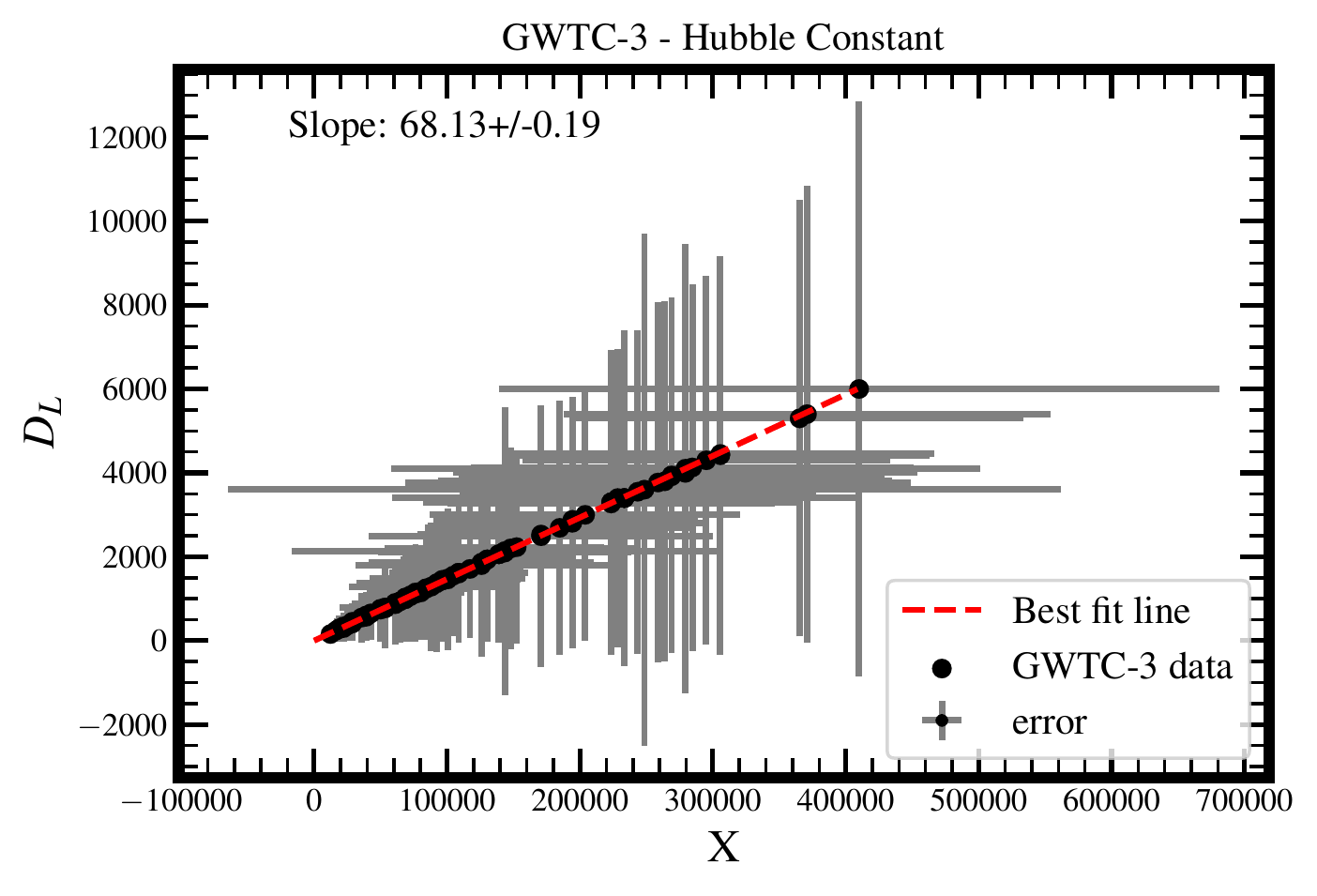}
     \caption{Hubble Parameter Estimation from the Pantheon sample, SCP Union 2.1 compilation and GWTC-3 dataset. Results are shown in the last column of table \ref{table2}}
\end{center}
\end{figure}

Equation \ref{DLeq}, can also be written as 
\begin{equation}
    d_L=\frac{X1}{H_0}
\end{equation}
where 
\begin{equation}
    X1= c(1+z) \int_{0}^{z} \frac{dz}{\Omega_{c h 0}^{1 / 2} X(z)}
\end{equation}
From the above equation, we inferred the value of the Hubble constant for all three observational samples, see Fig. \ref{H0estimate}. For the case of GWTC-3 dataset, errorbars are associated with both the variables, redshift and luminosity distance, hence we use orthogonal distance regression method for linear regression, by taking the average of the upper and lower error bar. The best fit is obtained between the observed and the theoretical value by minimizing
\begin{equation}
\sum_{i=1}^{n}\left[\frac{\left(y_{i}-\alpha-\beta X_{i}\right)}{\eta}+\left(x_{i}-X_{i}\right)\right]
\end{equation}
where $y_{i}$ is the observed value of distance modulus or the luminosity distance. $\alpha$-$\beta X_{i}$ is the value of distance modulus or luminosity distance obtained by VCG Model with applied constraints. $X_{i}$ is the absolute value of the other variable in study, redshift associated with the merger event and $x_{i}$ is the redshift value including the error range. $\eta$ is the ratio between the variance of the errorbars associated with the variables under study. By accounting for the variance in the error bars associated with the response variable, luminosity distance or the distance modulus of the event and the predictor variable, the redshift of the merger event, regression method is more sensitive towards the precision of both the predictor and the variables and minimizes the discrepancies brought in by the outliers in the dataset. 

However the Orthogonal Regression can be only used when the errors are normally distributed. In the case of the GWOSC dataset, the observed luminosity distance and therefore the distance modulus of the gravitational merger event is accompanied with non-Gaussian errors, therefore orthogonal regression can not provide accurate constraints on parameter $B_s$ and $n$. The non-Gaussian nature of the errorbars associated with the merger event makes Maximum Likelihood Estimation an ideal method to constraint the parameters under study.  

\subsection{Maximum Likelihood Estimation (MLE)}

The equation (\ref{chisqmodified}) which is for determining the $\chi^{2}$ goodness fit value has a dependency on the $H_0$, which is fixed. This results in substituting the Hubble parameter value from previous results to the equation. Therefore $\chi^{2}$ goodness value will be affected by any potential errors associated with the Hubble Parameter value from the previous results. This can not be avoided in the modified $\chi^{2}$ statistics analysis as explained in the previous section.\\
Maximum Likelihood Estimation (MLE) method can be utilized to infer the parameters $B_s$, $n$ and $H_0$ together, unlike modified $\chi^2$ statistics equation (\ref{chisqmodified}), where we obtained parameters $B_s$ and $n$, and was independent of $H_0$. This is extremely useful as even the $h_0$ parameter can also be determined using MLE along with $B_s$ and $n$, therefore the result becomes self-contained. The assumption that has to be taken is that the dataset are obtained from a same distribution family of the given parameters of the dataset and each individual data point on the dataset is not dependent on other data points to hold it's current value. This is a classic i.i.d. (Independent and Identically Distribution) assumption. The maximum likelihood estimation determines the combination of parameters by maximizing:
\begin{equation}
    \bigg(B_s,n,h_{0}\bigg)_{MLE} = \\ \underset{(B_s,n,h_{0})}{\arg\max}\\\mbox{\Large$\Pi$}f(a_i \| (B_s,n,h_{0}))
\label{ggg}
\end{equation}
here $\Pi f(a_i\|(B_s,n,h_{0})_{MLE}$ is a summation of probability function of every individual data point in the observed dataset. In general, the probability function for the total dataset under a given combination of parameters, $f(x_1,x_2,....x_n \| (B_s,n,h_{0}))$ is required in equation (\ref{ggg}). Since, the dataset follows i.i.d. distribution, the probability of combined data is equivalent to the summation of probability of each data point in the given dataset.
 \begin{figure*}[ht!]
    \centering
    \includegraphics[height=9cm, width=11cm]{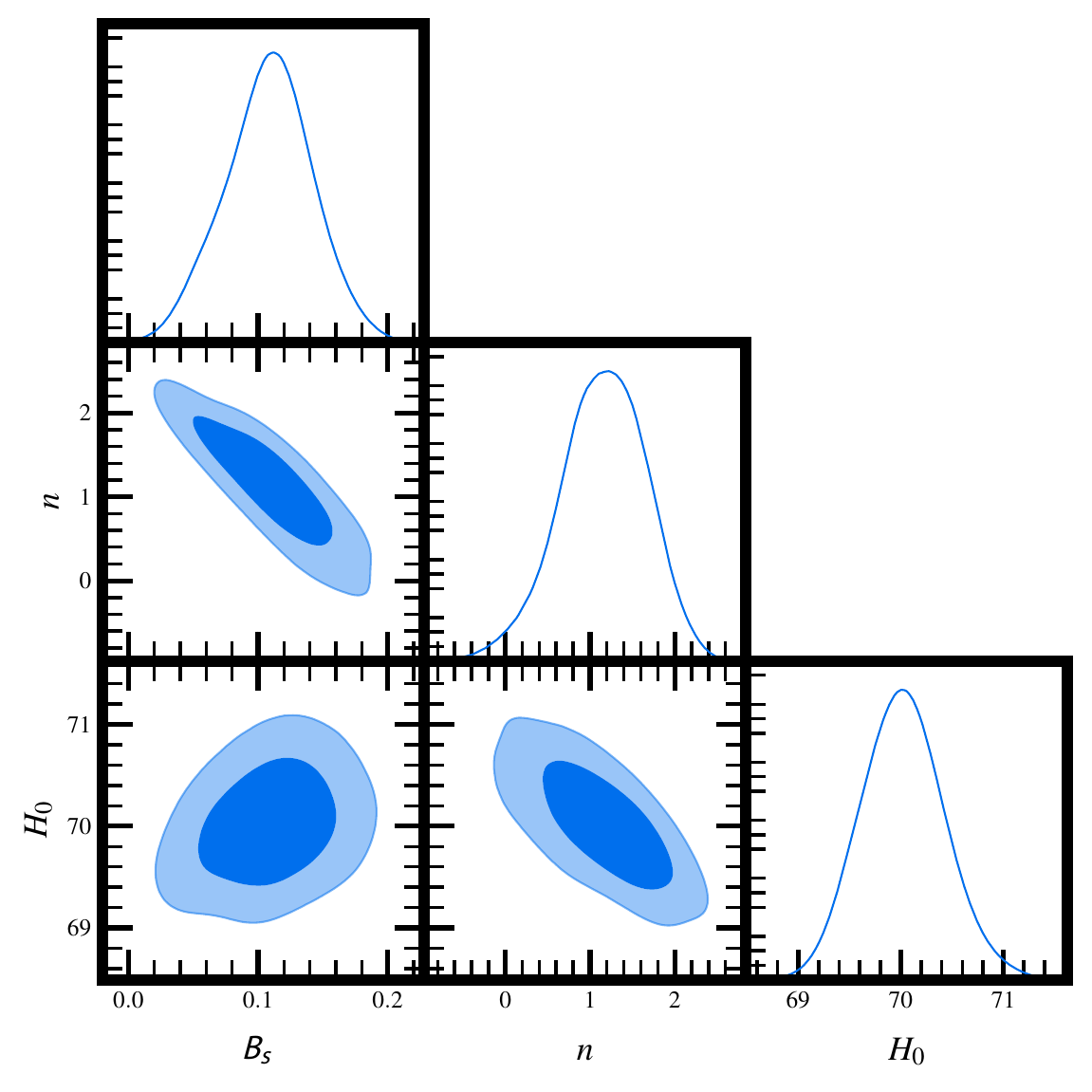}
    \caption{Cosmological parameters contours for the Pantheon sample using equation of distance modulus and Gaussian priors. Results are shown in table \ref{table4}.}
    \label{cobayapantheon}
\end{figure*}
\subsection{Maximum Likelihood Estimation using cobaya}
Cobaya (COde for BAYesian Analysis) (\citep{Torrado_2021,2019ascl.soft10019T}) is a coding framework built for statistical modelling to find arbitary posteriors from the given sets of parameters especially for Cosmology. cobaya determines Maximum Likelihood Estimation for the dataset from the given parameters using high advanced Monte Carlo samplers like MCMC from CosmoMC, nested samplers from Polychord or from a wide range of samplers incorporated in cobaya.
cobaya removes the dependency on the Hubble parameter value obtained from previous results and also significantly reduces the computing time to find the combination of the parameters $B_s$, $n$ and $H_{0}$ for which the probability for the given data distribution under MLE is maximum. The $\chi^{2}$ goodness fit algorithm takes much longer duration to find the combination of the parameters for which the $\chi^{2}$ goodness fit value is minimum in spite the fact that the parameters which are kept arbitrary are $B_s$ and $n$.

 \begin{figure*}[ht!]
    \centering
    \includegraphics[height=9cm, width=11cm]{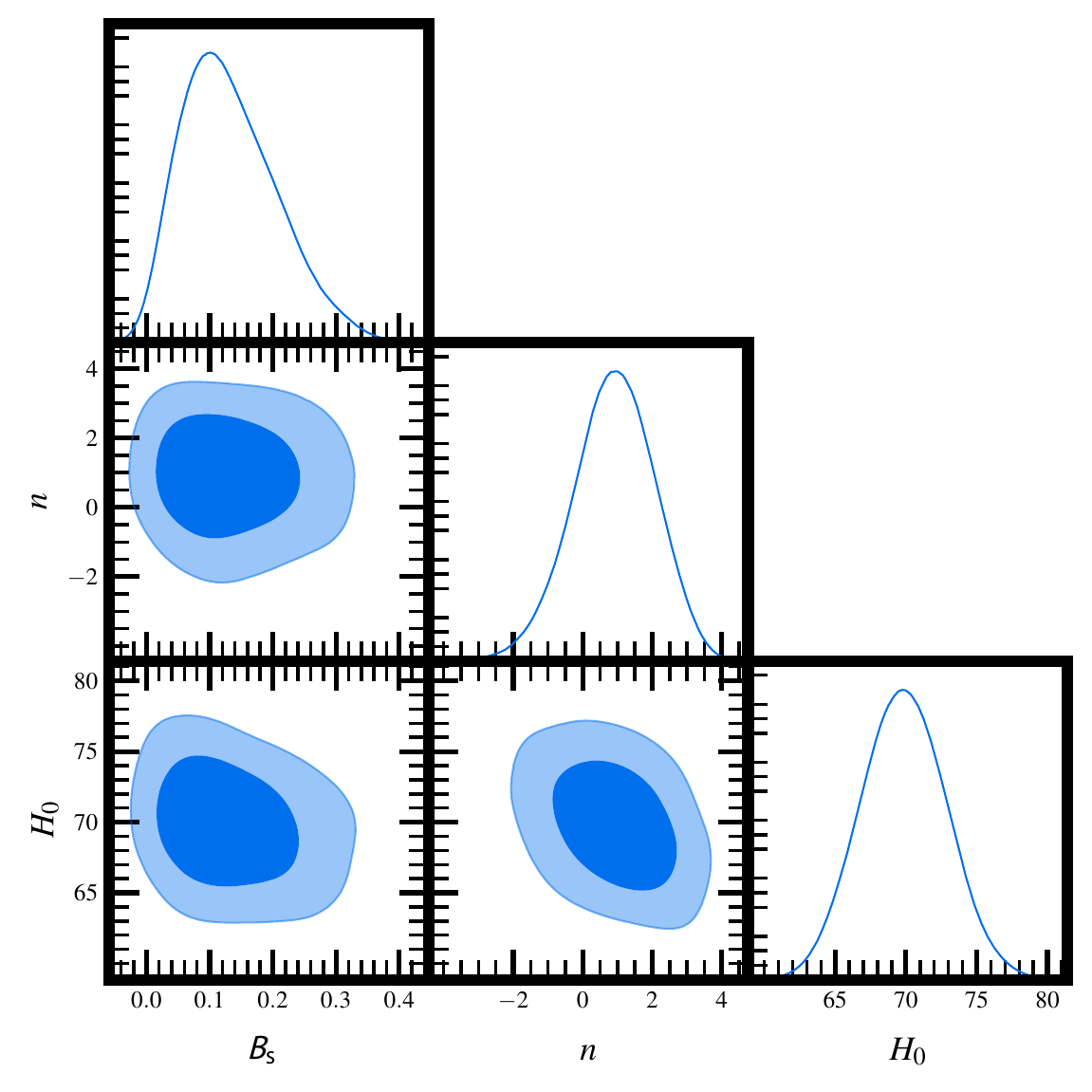}
    \caption{Cosmological parameters contours for the GWTC-3 sample using equation of distance modulus with Gaussian priors and taking average of assymmetrical error bars of GW results (Gaussian likelihood). Results are shown in table \ref{table1} }
    \label{cobayagaussianGW}
\end{figure*}
\begin{figure*}[ht!]
    \centering
    \includegraphics[height=9cm, width=11cm]{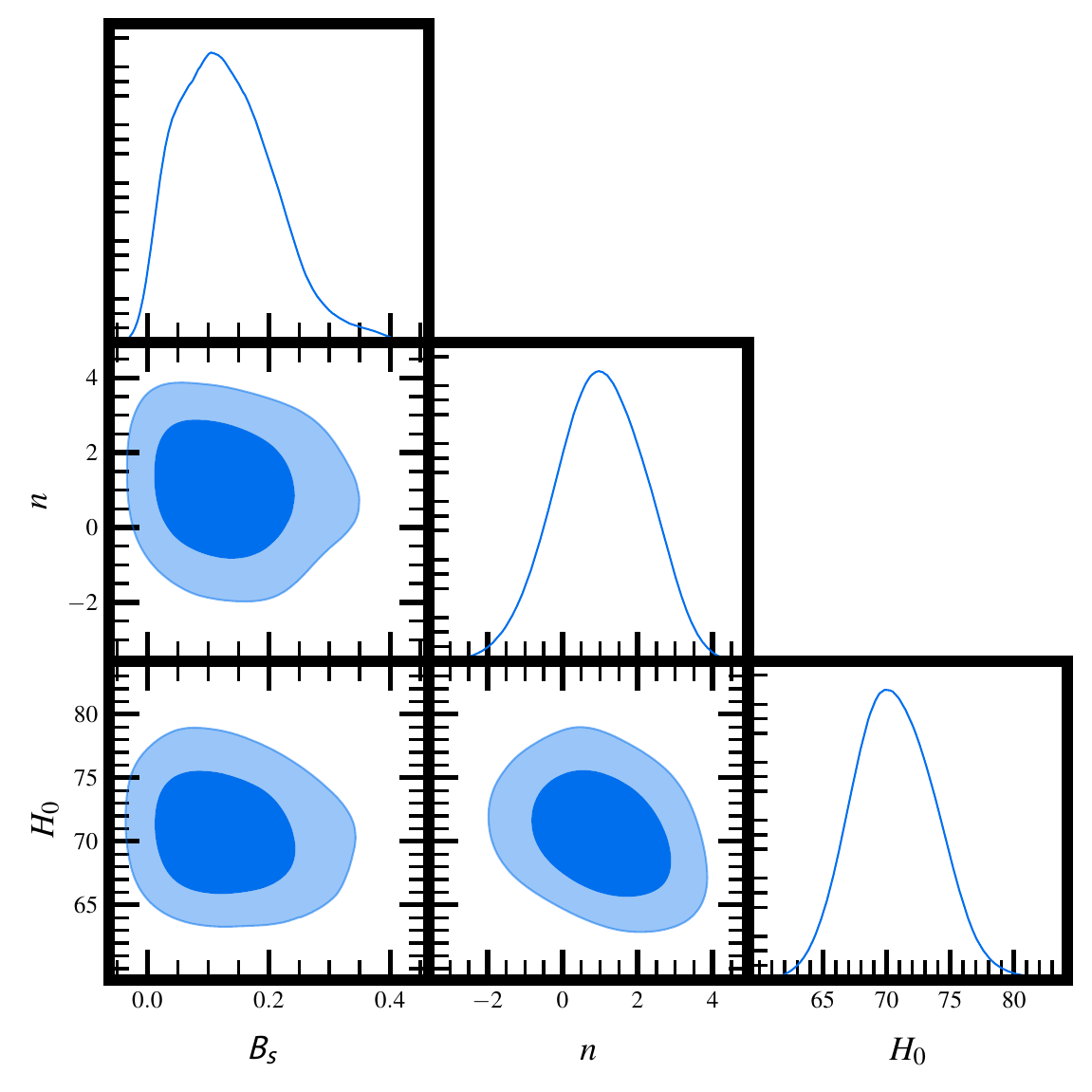}
    \caption{Cosmological parameters contours for the GWTC-3 sample using equation of distance modulus with Gaussian priors but taking modified likelihood because the dataset of GW has assymetrical error bars. Results are shown in table \ref{table1}}
    \label{cobayaassymmetryGW}
\end{figure*}
\begin{figure*}[ht!]
    \centering
    \includegraphics[height=9cm, width=11cm]{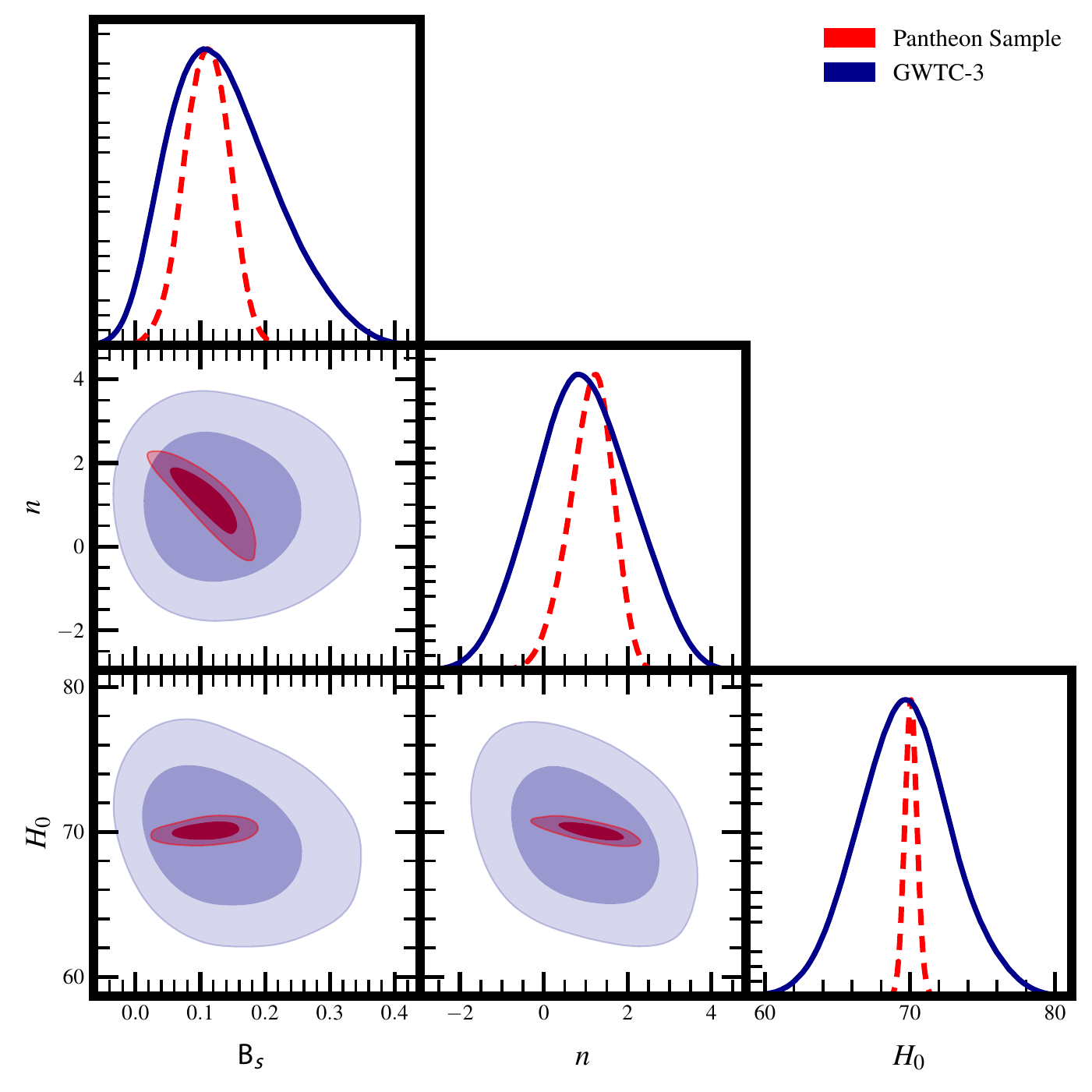}
    \caption{Cosmological parameters contours for the Pantheon sample and GWTC-3 together using equation of distance modulus, Gaussian priors and Gaussian likelihood. Results are shown in table \ref{table4}}
    \label{cobayapantheonGW}
\end{figure*}

\begin{table}[htb]
\resizebox{\columnwidth}{!}{%
\begin{tabular}{lccc}

\toprule

Likelihood & $B_s$ & n & $H_0$\\\midrule
Non-Gaussian & $0.130 \pm 0.079$ &  $1.025 \pm 1.120$ & $70.599 \pm 3.227$\\ \\

\vspace{3mm}

Gaussian & $0.130 \pm 0.076$ &  $0.897 \pm 1.182$ & $69.838 \pm 3.007$\\
\hline

\end{tabular}%
}
\caption{Constraints imposed on the parameters ($B_s$, $n$ \& $H_{0}$) of VCG Model from gravitational wave merger events using Gaussian and non-Gaussian likelihood functions. }
\label{table1}

\end{table}
From figures \ref{cobayagaussianGW} and \ref{cobayaassymmetryGW}, and table \ref{table1}, we can conclude that cosmological parameters values from Gaussian and non-Gaussian likelihoods fall within 1 sigma, hence, we would be using Gaussian likelihoods in further analysis for GWTC-3 dataset. We used Gaussian priors to calculate cosmological results for the Pantheon sample and GWTC-3 dataset as shown in figures \ref{cobayagaussianGW} and \ref{cobayapantheonGW}. For Gaussian priors, we considered mean value as the expectation value from the $\chi^2$ contour plots, see figure \ref{ContourPantheon} and \ref{chisqPantheonSCPGW}, and we doubled the standard deviation value $\sigma$ value which is then considered the new standard deviation for the MCMC. To find these best fitting parameters we use the D'Agostini \cite{d2005fits} Bayesian method, which takes into account the error bars on all the axes. We are not considering SCP Union 2.1 compilation here as the Pantheon sample has tighter constraints, as discussed before using $chi^2$ statistics. We have used modified likelihood function for the case of asymmetric errors in the GWTC-3 dataset. We are again able to show that the cosmological parameter values using the Pantheon sample and GWTC-3 dataset are consistent, and we are able to constrain the parameters using both datasets independently.

 \begin{figure*}[ht!]
    \centering
    \includegraphics[height=9cm, width=11cm]{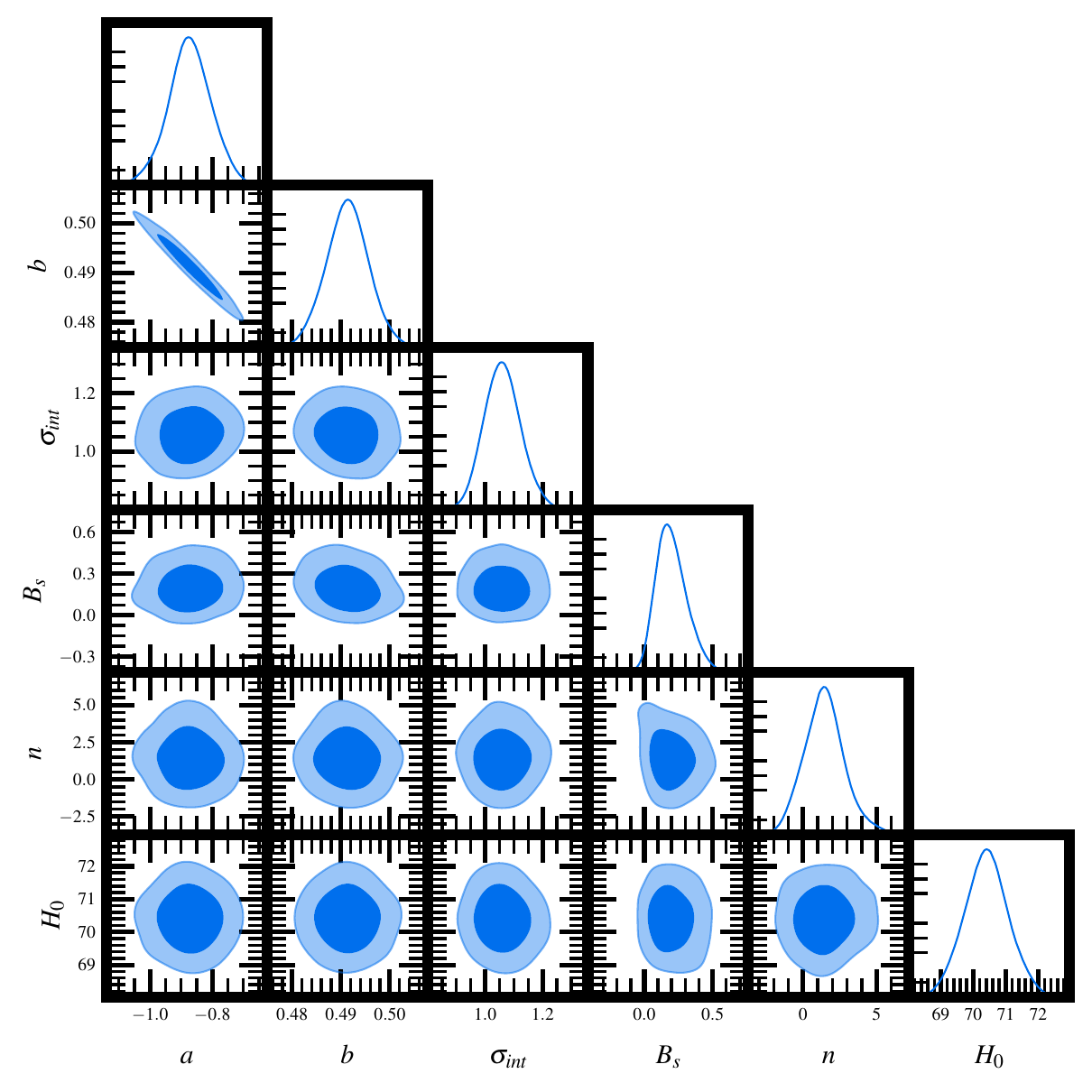}
    \caption{Cosmological parameters contours for the GRBs Platinum sample using equation of distance modulus and Gaussian priors. Results are shown in table \ref{table4}.}
    \label{cobayagrb}
\end{figure*}
\section{Gamma-ray Bursts 3D fundamental plane correlation as a cosmological
tool}\label{3D correlation}
Gamma-ray bursts (GRBs) are incredibly powerful phenomena. They are the brightest objects after the Big Bang, and some of the farthest astrophysical objects ever detected \citep{Paczynski1986,Planck2018,Kumar2015}. These features make GRBs promising cosmological tools, similar to supernovae type Ia (SNe Ia). GRBs are extremely luminous, which allows them to be observed at very large distances, corresponding to high redshifts. Indeed, GRBs have been observed up to redshifts of 8.2 and 9.4 \citep{tanvir2009,cucchiara2011}, while SNe Ia have only been observed up to redshift 2.26 \citep{Rodney2015}. However, using GRBs as cosmological tools requires a full understanding of their physical mechanisms. Both their energy emission mechanisms and progenitors are still being studied by the scientific community.

There are two main scenarios for the birth of GRBs. The first scenario is the explosion of a very massive star at the end of its lifetime \citep{Narayan1992,Woosley1993,MacFadyen1999,Nagataki2007, Nagataki2009}. This is followed by a core-collapse SNe \citep{Stanek2003,MacFadyen2001}. The second scenario is the coalescence of two compact objects, like black holes (BHs) or neutron stars (NSs) \citep{Lattimer1976,Eichler1989,Li1998,Rowlinson2014,Rea2015}. The most probable frameworks for the central engine that powers the GRB consider the following astrophysical objects: BHs, NSs, or fast spinning newly born highly magnetized NSs magnetars \citep{Usov1992,Liang2018,Ai2018}. GRBs can be divided into two main categories: short and long. Short GRBs are associated with the merging of two compact objects, while long GRBs are associated with the core collapse of a very massive star. GRB light curves (LCs) can be divided into two main phases: the prompt emission and the afterglow. The prompt emission is a rapid burst of gamma rays, while the afterglow is a longer-lasting emission that can be detected in X-ray, optical, and radio wavelengths. The plateau phase of GRBs is a flat part of the LC that follows the prompt emission. It was discovered by the Neil Gehrels Swift Observatory (Swift) and typically ranges from $10^2$ to $10^5$ seconds in duration. The plateau phase is thought to be caused by the spin-down of a newly born magnetar or the external shock model.

In the past decades, many efforts have been made to find possible correlations between physical features of GRBs. One of these correlations is the so-called Dainotti relation, which links the time at the end of the plateau emission measured in the rest frame, $T^*_X$, with the corresponding X-ray luminosity of the LC, $L_X$ \citep{Dainotti2008}. This correlation is theoretically supported by the magnetar model. Its extension in three dimensions has been discovered by adding the prompt peak luminosity, $L_{\text{peak}}$, and is known as the fundamental plane correlation or the 3D Dainotti relation \citep{Dainotti2016, Dainotti2017c}. It has also shown that GRBs can be used as cosmological probes. \noindent The fundamental plane relation has the following form:
\begin{linenomath*}
\begin{equation}
\log L_X = c + a\cdot \log T^{*}_{X} + b \cdot( \log L_{peak}),
\label{isotropic}
\end{equation} 
\end{linenomath*}
\noindent where $a$ and $b$ are the best-fit parameters given by the D’Agostini procedure \cite{Dago05} linked to $\log T^{*}_{X}$ and $\log L_{peak}$, respectively, while $c$ is the normalization.
In this study, we follow the same approach as in \citep{dainotti2023gamma} and use the Platinum Sample of GRBs as a cosmological probe. The Platinum Sample of GRBs is a compilation of the highest-quality GRBs with well-measured redshifts. The GRBs in this sample are well-suited for cosmological studies because they are bright and can be seen to high redshifts.  We further combine the GRB Platinum sample with the Supernovae Ia Pantheon sample and GWTC-3 Catalog using a Bayesian framework, and compute the cosmological parameters. This facilitates the examination of whether the inclusion of GRBs in the analysis would corroborate the findings observed when studying individual samples independently, as well as the extent to which it could augment the accuracy of the cosmological parameters.

\section{Analysis and Result}\label{sec5}

We used $\chi^2$  minimization to obtain the parameters $B_s$ and n of VCG model using the Pantheon sample, SCP Union 2.1 compilation and through MLE for GWTC-3 dataset. Subsequently, we used the obtained parameters to infer the Hubble constant. Values of these parameters can be found in table \ref{table2}. For comparison and completeness, we also tested our $\chi^2$ program with the Pantheon sample and obtained for Standard Model of Cosmology, $\Omega_m = 0.298 \pm 0.0087$ and $H_0 = 71.985 \pm 0.663$.

 \begin{figure*}[ht!]
    \centering
    \includegraphics[height=9cm, width=11cm]{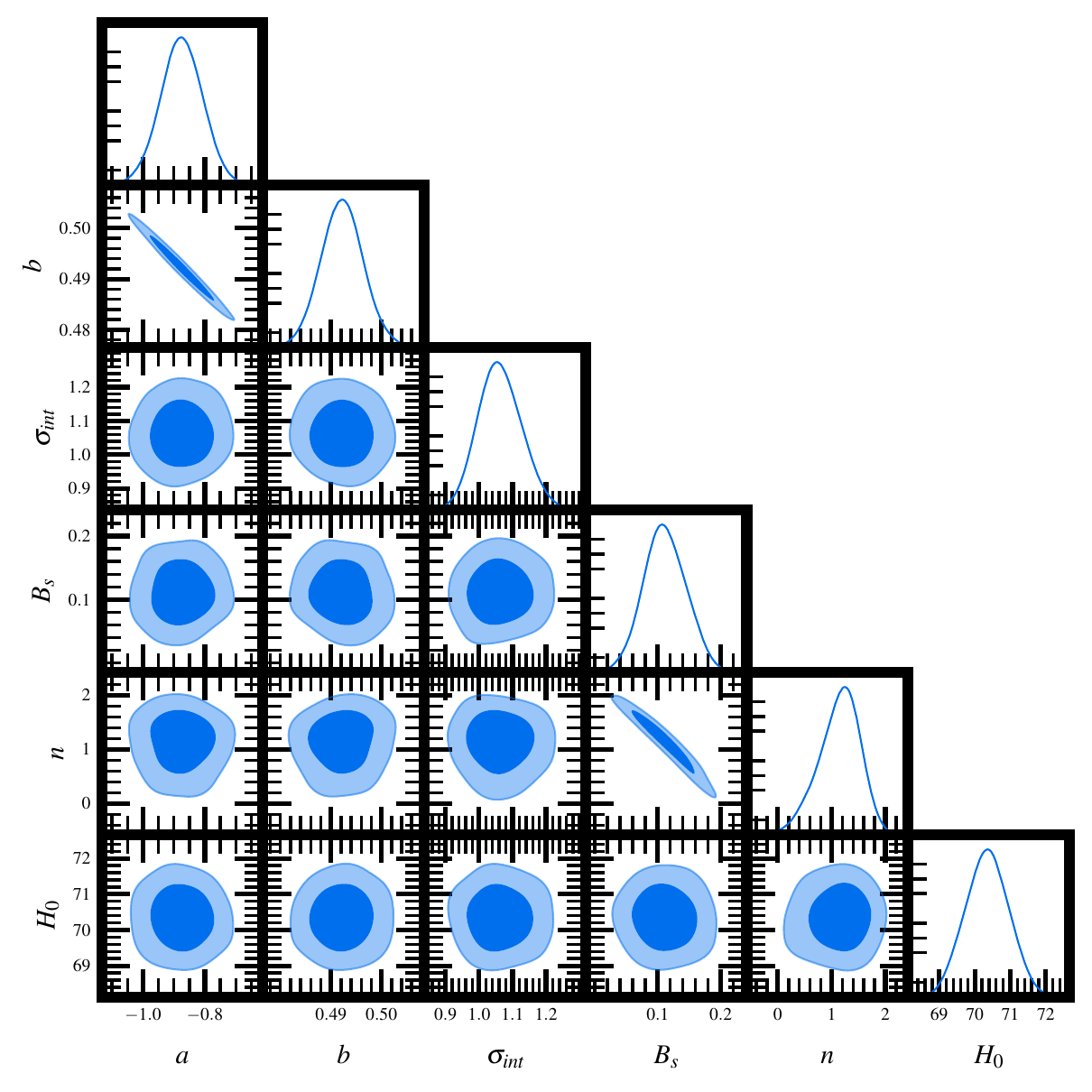}
    \caption{Cosmological parameters contours for the combined samples (SNIe Pantheon+ GRBs Platinum + GWTC-3) using equation of distance modulus and Gaussian priors. Results are shown in table \ref{table4}.}
    \label{cobayacombined}
\end{figure*}

\begin{table}[htb]
\resizebox{\columnwidth}{!}{%
\begin{tabular}{lccc}
\toprule
Dataset & $B_s$ & n & $H_0$\\\midrule
Pantheon &  $0.11^{+0.024,+0.043}_{-0.024,-0.037}$ &  $1.260^{+0.205,+0.410}_{-0.308,-0.513}$ & $70.47^{+0.13}_{-0.13}$\\ \\

\vspace{3mm}

SCP Union 2.1 &  $0.096^{+0.049,+0.073}_{-0.043,-0.062}$ &  $1.301^{+0.485,+0.623}_{-0.692,-0.969}$ & $69.41^{+0.39}_{-0.39}$\\ \\

\vspace{3mm}

GWTC-3 &$0.213^{+0.355, +0.413}_{-0.203, -0.203}$ & $-1.744^{+4.170, 4.667}_{-98.265, -98.265}$ & $68.13^{+0.19}_{-0.19}$\\
\hline
\end{tabular}%
}
\caption{Constraints imposed on the parameters ($B_s$, $n$ \& $H_{0}$) of VCG Model from Supernovae Events and gravitational wave merger events using modified $\chi^2$ statistics governed by equation \ref{chisqmodified}. }
\label{table2}
\end{table}

From figure \ref{PS Plot}, we can conclude that VCG model is a viable model which is capable of explaining the observed expansion of the universe using the Pantheon sample, SCP Union 2.1 compilation and GWTC-3 dataset. As the GWTC-3 dataset has asymmetrical error bars, we can't use $\chi^2$  minimization method to obtain the VCG model parameters for GWTC-3 dataset. Hence, we used cobaya for MCMC computation using modified likelihood function and the Gaussian priors. We used results obtained from $\chi^2$ in table \ref{table2} using Pantheon sample as the priors and we let all the parameters vary simultaneously. We performed the analysis using SNIe Pantheon sample, GWTC-3 catalog, GRBs Platinum sample and combining all three. We find the best fit results of the constants in the fundamental plane relation are: $ a =-0.88 \pm 0.06$,
$b =0.49 \pm 0.01$, and $\sigma_{int}=1.06 \pm 0.06$, where $\sigma_{int}$ is the intrinsic scatter of the correlation. The results are shown in table \ref{table4}.
To complete our analysis in table \ref{Table3}, we present comparison of our results with the VGCG and the GCG models also. 

\begin{table}[ht]
\centering
\resizebox{\columnwidth}{!}{%
\begin{tabular}{lcccccc}
\toprule
Dataset & $\chi^{2}_{min}$/$dof$ & $h$ & $\Omega_{b0}h^2$ & $B_s$ & $\alpha$ & $\xi_0$\\
\midrule
\multicolumn{7}{c}{VGCG}\\
\midrule \\
SnIa & 1036.4/1043 & $0.70^{+0.18}_{-0.20}$ & $0.02242^{+0.00014}_{-0.00014}$ & $0.50^{+0.18}_{-0.21}$ & $0.93^{+0.69}_{-0.61}$ & $0.15^{+0.09}_{-0.09}$\\ \\
\midrule
\multicolumn{7}{c}{GCG} \\
\midrule\\
SnIa & 1036.3/1044 & $0.71^{+0.20}_{-0.18}$ & $0.02242^{+0.00014}_{-0.00014}$ & $0.96^{+0.05}_{-0.04}$ & $1.13^{+0.44}_{-0.28}$ & - \\ \\
\hline
\end{tabular}%
}
\caption{Results of constraining the cosmological parameters with Pantheon dataset in other dissipative fluids models of cosmology like VGCG and GCG \cite{HER-AL}}

\label{Table3}
\end{table}

\begin{table}[htb]
\centering

\resizebox{\columnwidth}{!}{%

\begin{tabular}{lccc}
\toprule
Dataset & $B_s$ & n & $H_0$\\\midrule

Pantheon &  $0.11 \pm 0.03$ &  $1.16 \pm 0.51$ & $70.02 \pm 0.41$\\ \\

\vspace{3mm}

GWTC-3 &  $0.13 \pm 0.08$ &  $0.90 \pm 1.18$ & $69.84 \pm 3.01$\\

\vspace{3mm}

GRBs &  $0.20 \pm 0.11$ &  $1.45 \pm 1.40$ & $70.41 \pm 0.67$\\

\vspace{3mm}

Combined &  $0.11 \pm 0.03$ &  $1.14 \pm 0.36$ & $70.34 \pm 0.61$\\

\hline

\end{tabular}%
}
\caption{Constraints imposed on the parameters ($B_s$, $n$ \& $H_{0}$) of VCG Model from Supernovae events, GRB events, GW merger events and combined of all three datasets using distance modulus equation, Gaussian priors and Gaussian likelihood.}
\label{table4}
\end{table}

\section{Conclusions}\label{sec6}

The Variable Chaplygin gas model is indeed able to explain the evolution of the universe in account for the gravitational waves from merger events. As the VCG Model speculate, the Chaplygin gas is expected to behave like a non-relativistic entity and later evolves to account for the accelerated expansion observed in the current epoch of the Universe. The best fit obtained from the VCG model with the GWTC-3 Dataset lies within confidence levels obtained in analysis conducted by \cite{Sethi_2006} and \cite{VCGPrime} and also is compatible with the constraints obtained from GRBs. The GWTC-3 dataset has also brought in further constraints on the parameters $B_s$ and $n$. The constraints obtained using GWTC-3 are expected to become  better as the dataset becomes bigger with more detections. These results are indeed limited to the fact that locating sources of Gravitational waves is still subject to improvement in the detection techniques.

\section{Acknowledgement}\label{sec7}

We would acknowledge Isophote for providing us the opportunity to work in this project. Dr. Geetanjali Sethi is also thankful to Principal, St. Stephen's College for allowing her to be a part of this project. Ashley Chraya is thankful to Department of Physics, Indian Institute of Science Education and Research, Mohali for enabling him to be a part of this project. We also thank Dr. Akshay Rana, St. Stephen's College for his inputs and suggestions.

\section*{Declarations}
\subsection*{Funding and/or Conflicts of interests / Competing interests}
The authors declare no conflict of interest.





\bibliography{main_paper}


\begin{thebibliography}{77}
\ifx \bisbn   \undefined \def \bisbn  #1{ISBN #1}\fi
\ifx \binits  \undefined \def \binits#1{#1}\fi
\ifx \bauthor  \undefined \def \bauthor#1{#1}\fi
\ifx \batitle  \undefined \def \batitle#1{#1}\fi
\ifx \bjtitle  \undefined \def \bjtitle#1{#1}\fi
\ifx \bvolume  \undefined \def \bvolume#1{\textbf{#1}}\fi
\ifx \byear  \undefined \def \byear#1{#1}\fi
\ifx \bissue  \undefined \def \bissue#1{#1}\fi
\ifx \bfpage  \undefined \def \bfpage#1{#1}\fi
\ifx \blpage  \undefined \def \blpage #1{#1}\fi
\ifx \burl  \undefined \def \burl#1{\textsf{#1}}\fi
\ifx \doiurl  \undefined \def \doiurl#1{\url{https://doi.org/#1}}\fi
\ifx \betal  \undefined \def \betal{\textit{et al.}}\fi
\ifx \binstitute  \undefined \def \binstitute#1{#1}\fi
\ifx \binstitutionaled  \undefined \def \binstitutionaled#1{#1}\fi
\ifx \bctitle  \undefined \def \bctitle#1{#1}\fi
\ifx \beditor  \undefined \def \beditor#1{#1}\fi
\ifx \bpublisher  \undefined \def \bpublisher#1{#1}\fi
\ifx \bbtitle  \undefined \def \bbtitle#1{#1}\fi
\ifx \bedition  \undefined \def \bedition#1{#1}\fi
\ifx \bseriesno  \undefined \def \bseriesno#1{#1}\fi
\ifx \blocation  \undefined \def \blocation#1{#1}\fi
\ifx \bsertitle  \undefined \def \bsertitle#1{#1}\fi
\ifx \bsnm \undefined \def \bsnm#1{#1}\fi
\ifx \bsuffix \undefined \def \bsuffix#1{#1}\fi
\ifx \bparticle \undefined \def \bparticle#1{#1}\fi
\ifx \barticle \undefined \def \barticle#1{#1}\fi
\bibcommenthead
\ifx \bconfdate \undefined \def \bconfdate #1{#1}\fi
\ifx \botherref \undefined \def \botherref #1{#1}\fi
\ifx \url \undefined \def \url#1{\textsf{#1}}\fi
\ifx \bchapter \undefined \def \bchapter#1{#1}\fi
\ifx \bbook \undefined \def \bbook#1{#1}\fi
\ifx \bcomment \undefined \def \bcomment#1{#1}\fi
\ifx \oauthor \undefined \def \oauthor#1{#1}\fi
\ifx \citeauthoryear \undefined \def \citeauthoryear#1{#1}\fi
\ifx \endbibitem  \undefined \def \endbibitem {}\fi
\ifx \bconflocation  \undefined \def \bconflocation#1{#1}\fi
\ifx \arxivurl  \undefined \def \arxivurl#1{\textsf{#1}}\fi
\csname PreBibitemsHook\endcsname

\bibitem{Perlmutter_1999}
\begin{barticle}
\bauthor{\bsnm{Perlmutter}, \binits{S.}},
\bauthor{\bsnm{Aldering}, \binits{G.}},
\bauthor{\bsnm{Goldhaber}, \binits{G.}},
\bauthor{\bsnm{Knop}, \binits{R.A.}},
\bauthor{\bsnm{Nugent}, \binits{P.}},
\bauthor{\bsnm{Castro}, \binits{P.G.}},
\bauthor{\bsnm{Deustua}, \binits{S.}},
\bauthor{\bsnm{Fabbro}, \binits{S.}},
\bauthor{\bsnm{Goobar}, \binits{A.}},
\bauthor{\bsnm{Groom}, \binits{D.E.}},
\bauthor{\bsnm{Hook}, \binits{I.M.}},
\bauthor{\bsnm{Kim}, \binits{A.G.}},
\bauthor{\bsnm{Kim}, \binits{M.Y.}},
\bauthor{\bsnm{Lee}, \binits{J.C.}},
\bauthor{\bsnm{Nunes}, \binits{N.J.}},
\bauthor{\bsnm{Pain}, \binits{R.}},
\bauthor{\bsnm{Pennypacker}, \binits{C.R.}},
\bauthor{\bsnm{Quimby}, \binits{R.}},
\bauthor{\bsnm{Lidman}, \binits{C.}},
\bauthor{\bsnm{Ellis}, \binits{R.S.}},
\bauthor{\bsnm{Irwin}, \binits{M.}},
\bauthor{\bsnm{McMahon}, \binits{R.G.}},
\bauthor{\bsnm{Ruiz-Lapuente}, \binits{P.}},
\bauthor{\bsnm{Walton}, \binits{N.}},
\bauthor{\bsnm{Schaefer}, \binits{B.}},
\bauthor{\bsnm{Boyle}, \binits{B.J.}},
\bauthor{\bsnm{Filippenko}, \binits{A.V.}},
\bauthor{\bsnm{Matheson}, \binits{T.}},
\bauthor{\bsnm{Fruchter}, \binits{A.S.}},
\bauthor{\bsnm{Panagia}, \binits{N.}},
\bauthor{\bsnm{Newberg}, \binits{H.J.M.}},
\bauthor{\bsnm{Couch}, \binits{W.J.}},
\bauthor{\bsnm{Project}, \binits{T.S.C.}}:
\batitle{Measurements of $\omega$ and $\lambda$ from 42 high-redshift
  supernovae}.
\bjtitle{The Astrophysical Journal}
\bvolume{517}(\bissue{2}),
\bfpage{565}--\blpage{586}
(\byear{1999}).
\doiurl{10.1086/307221}
\end{barticle}
\endbibitem

\bibitem{riess1998observational}
\begin{barticle}
\bauthor{\bsnm{Riess}, \binits{A.G.}},
\bauthor{\bsnm{Filippenko}, \binits{A.V.}},
\bauthor{\bsnm{Challis}, \binits{P.}},
\bauthor{\bsnm{Clocchiatti}, \binits{A.}},
\bauthor{\bsnm{Diercks}, \binits{A.}},
\bauthor{\bsnm{Garnavich}, \binits{P.M.}},
\bauthor{\bsnm{Gilliland}, \binits{R.L.}},
\bauthor{\bsnm{Hogan}, \binits{C.J.}},
\bauthor{\bsnm{Jha}, \binits{S.}},
\bauthor{\bsnm{Kirshner}, \binits{R.P.}}, \betal:
\batitle{Observational evidence from supernovae for an accelerating universe
  and a cosmological constant}.
\bjtitle{The Astronomical Journal}
\bvolume{116}(\bissue{3}),
\bfpage{1009}
(\byear{1998})
\end{barticle}
\endbibitem

\bibitem{Spergel_2003}
\begin{barticle}
\bauthor{\bsnm{Spergel}, \binits{D.N.}},
\bauthor{\bsnm{Verde}, \binits{L.}},
\bauthor{\bsnm{Peiris}, \binits{H.V.}},
\bauthor{\bsnm{Komatsu}, \binits{E.}},
\bauthor{\bsnm{Nolta}, \binits{M.R.}},
\bauthor{\bsnm{Bennett}, \binits{C.L.}},
\bauthor{\bsnm{Halpern}, \binits{M.}},
\bauthor{\bsnm{Hinshaw}, \binits{G.}},
\bauthor{\bsnm{Jarosik}, \binits{N.}},
\bauthor{\bsnm{Kogut}, \binits{A.}},
\bauthor{\bsnm{Limon}, \binits{M.}},
\bauthor{\bsnm{Meyer}, \binits{S.S.}},
\bauthor{\bsnm{Page}, \binits{L.}},
\bauthor{\bsnm{Tucker}, \binits{G.S.}},
\bauthor{\bsnm{Weiland}, \binits{J.L.}},
\bauthor{\bsnm{Wollack}, \binits{E.}},
\bauthor{\bsnm{Wright}, \binits{E.L.}}:
\batitle{First-year wilkinson microwave anisotropy probe ( {WMAP} )
  observations: Determination of cosmological parameters}.
\bjtitle{The Astrophysical Journal Supplement Series}
\bvolume{148}(\bissue{1}),
\bfpage{175}--\blpage{194}
(\byear{2003}).
\doiurl{10.1086/377226}
\end{barticle}
\endbibitem

\bibitem{miller1999measurement}
\begin{barticle}
\bauthor{\bsnm{Miller}, \binits{A.D.}},
\bauthor{\bsnm{Caldwell}, \binits{R.}},
\bauthor{\bsnm{Devlin}, \binits{M.J.}},
\bauthor{\bsnm{Dorwart}, \binits{W.}},
\bauthor{\bsnm{Herbig}, \binits{T.}},
\bauthor{\bsnm{Nolta}, \binits{M.}},
\bauthor{\bsnm{Page}, \binits{L.}},
\bauthor{\bsnm{Puchalla}, \binits{J.}},
\bauthor{\bsnm{Torbet}, \binits{E.}},
\bauthor{\bsnm{Tran}, \binits{H.}}:
\batitle{A measurement of the angular power spectrum of the cosmic microwave
  background from l= 100 to 400}.
\bjtitle{The Astrophysical Journal}
\bvolume{524}(\bissue{1}),
\bfpage{1}
(\byear{1999})
\end{barticle}
\endbibitem

\bibitem{bahcall1999cosmic}
\begin{barticle}
\bauthor{\bsnm{Bahcall}, \binits{N.A.}},
\bauthor{\bsnm{Ostriker}, \binits{J.P.}},
\bauthor{\bsnm{Perlmutter}, \binits{S.}},
\bauthor{\bsnm{Steinhardt}, \binits{P.J.}}:
\batitle{The cosmic triangle: Revealing the state of the universe}.
\bjtitle{science}
\bvolume{284}(\bissue{5419}),
\bfpage{1481}--\blpage{1488}
(\byear{1999})
\end{barticle}
\endbibitem

\bibitem{PhysRevD.37.3406}
\begin{barticle}
\bauthor{\bsnm{Ratra}, \binits{B.}},
\bauthor{\bsnm{Peebles}, \binits{P.J.E.}}:
\batitle{Cosmological consequences of a rolling homogeneous scalar field}.
\bjtitle{Phys. Rev. D}
\bvolume{37},
\bfpage{3406}--\blpage{3427}
(\byear{1988}).
\doiurl{10.1103/PhysRevD.37.3406}
\end{barticle}
\endbibitem

\bibitem{PhysRevLett.82.896}
\begin{barticle}
\bauthor{\bsnm{Zlatev}, \binits{I.}},
\bauthor{\bsnm{Wang}, \binits{L.}},
\bauthor{\bsnm{Steinhardt}, \binits{P.J.}}:
\batitle{Quintessence, cosmic coincidence, and the cosmological constant}.
\bjtitle{Phys. Rev. Lett.}
\bvolume{82},
\bfpage{896}--\blpage{899}
(\byear{1999}).
\doiurl{10.1103/PhysRevLett.82.896}
\end{barticle}
\endbibitem

\bibitem{guo2007parametrizations}
\begin{barticle}
\bauthor{\bsnm{Guo}, \binits{Z.-K.}},
\bauthor{\bsnm{Ohta}, \binits{N.}},
\bauthor{\bsnm{Zhang}, \binits{Y.-Z.}}:
\batitle{Parametrizations of the dark energy density and scalar potentials}.
\bjtitle{Modern Physics Letters A}
\bvolume{22}(\bissue{12}),
\bfpage{883}--\blpage{890}
(\byear{2007})
\end{barticle}
\endbibitem

\bibitem{bento2002two}
\begin{barticle}
\bauthor{\bsnm{Bento}, \binits{M.}},
\bauthor{\bsnm{Bertolami}, \binits{O.}},
\bauthor{\bsnm{Santos}, \binits{N.}}:
\batitle{A two-field quintessence model}.
\bjtitle{Physical Review D}
\bvolume{65}(\bissue{6}),
\bfpage{067301}
(\byear{2002})
\end{barticle}
\endbibitem

\bibitem{sahni2000case}
\begin{barticle}
\bauthor{\bsnm{Sahni}, \binits{V.}},
\bauthor{\bsnm{Starobinsky}, \binits{A.}}:
\batitle{The case for a positive cosmological $\lambda$-term}.
\bjtitle{International Journal of Modern Physics D}
\bvolume{9}(\bissue{04}),
\bfpage{373}--\blpage{443}
(\byear{2000})
\end{barticle}
\endbibitem

\bibitem{padmanabhan2003cosmological}
\begin{barticle}
\bauthor{\bsnm{Padmanabhan}, \binits{T.}}:
\batitle{Cosmological constant—the weight of the vacuum}.
\bjtitle{Physics Reports}
\bvolume{380}(\bissue{5-6}),
\bfpage{235}--\blpage{320}
(\byear{2003})
\end{barticle}
\endbibitem

\bibitem{RevModPhys.75.559}
\begin{barticle}
\bauthor{\bsnm{Peebles}, \binits{P.J.E.}},
\bauthor{\bsnm{Ratra}, \binits{B.}}:
\batitle{The cosmological constant and dark energy}.
\bjtitle{Rev. Mod. Phys.}
\bvolume{75},
\bfpage{559}--\blpage{606}
(\byear{2003}).
\doiurl{10.1103/RevModPhys.75.559}
\end{barticle}
\endbibitem

\bibitem{RevModPhys.61.1}
\begin{barticle}
\bauthor{\bsnm{Weinberg}, \binits{S.}}:
\batitle{The cosmological constant problem}.
\bjtitle{Rev. Mod. Phys.}
\bvolume{61},
\bfpage{1}--\blpage{23}
(\byear{1989}).
\doiurl{10.1103/RevModPhys.61.1}
\end{barticle}
\endbibitem

\bibitem{barrow2006cosmologies}
\begin{barticle}
\bauthor{\bsnm{Barrow}, \binits{J.D.}},
\bauthor{\bsnm{Clifton}, \binits{T.}}:
\batitle{Cosmologies with energy exchange}.
\bjtitle{Physical Review D}
\bvolume{73}(\bissue{10}),
\bfpage{103520}
(\byear{2006})
\end{barticle}
\endbibitem

\bibitem{gonzalez2006dynamics}
\begin{barticle}
\bauthor{\bsnm{Gonzalez}, \binits{T.}},
\bauthor{\bsnm{Leon}, \binits{G.}},
\bauthor{\bsnm{Quiros}, \binits{I.}}:
\batitle{Dynamics of quintessence models of dark energy with exponential
  coupling to dark matter}.
\bjtitle{Classical and Quantum Gravity}
\bvolume{23}(\bissue{9}),
\bfpage{3165}
(\byear{2006})
\end{barticle}
\endbibitem

\bibitem{boehmer2008dynamics}
\begin{barticle}
\bauthor{\bsnm{Boehmer}, \binits{C.G.}},
\bauthor{\bsnm{Caldera-Cabral}, \binits{G.}},
\bauthor{\bsnm{Lazkoz}, \binits{R.}},
\bauthor{\bsnm{Maartens}, \binits{R.}}:
\batitle{Dynamics of dark energy with a coupling to dark matter}.
\bjtitle{Physical Review D}
\bvolume{78}(\bissue{2}),
\bfpage{023505}
(\byear{2008})
\end{barticle}
\endbibitem

\bibitem{jamil2009constraining}
\begin{barticle}
\bauthor{\bsnm{Jamil}, \binits{M.}},
\bauthor{\bsnm{Rashid}, \binits{M.A.}}:
\batitle{Constraining the coupling constant between dark energy and dark
  matter}.
\bjtitle{The European Physical Journal C}
\bvolume{60}(\bissue{1}),
\bfpage{141}--\blpage{147}
(\byear{2009})
\end{barticle}
\endbibitem

\bibitem{kamenshchik2001alternative}
\begin{barticle}
\bauthor{\bsnm{Kamenshchik}, \binits{A.}},
\bauthor{\bsnm{Moschella}, \binits{U.}},
\bauthor{\bsnm{Pasquier}, \binits{V.}}:
\batitle{An alternative to quintessence}.
\bjtitle{Physics Letters B}
\bvolume{511}(\bissue{2-4}),
\bfpage{265}--\blpage{268}
(\byear{2001})
\end{barticle}
\endbibitem

\bibitem{bilic2002unification}
\begin{barticle}
\bauthor{\bsnm{Bili{\'c}}, \binits{N.}},
\bauthor{\bsnm{Tupper}, \binits{G.B.}},
\bauthor{\bsnm{Viollier}, \binits{R.D.}}:
\batitle{Unification of dark matter and dark energy: the inhomogeneous
  chaplygin gas}.
\bjtitle{Physics Letters B}
\bvolume{535}(\bissue{1-4}),
\bfpage{17}--\blpage{21}
(\byear{2002})
\end{barticle}
\endbibitem

\bibitem{saadat2014vol}
\begin{botherref}
\oauthor{\bsnm{Saadat}, \binits{H.}},
\oauthor{\bsnm{Pourhassan}, \binits{B.}}:
vol. 53.
Int. J. Theor. Phys,
1168
(2014)
\end{botherref}
\endbibitem

\bibitem{debnath2004role}
\begin{barticle}
\bauthor{\bsnm{Debnath}, \binits{U.}},
\bauthor{\bsnm{Banerjee}, \binits{A.}},
\bauthor{\bsnm{Chakraborty}, \binits{S.}}:
\batitle{Role of modified chaplygin gas in accelerated universe}.
\bjtitle{Classical and Quantum Gravity}
\bvolume{21}(\bissue{23}),
\bfpage{5609}
(\byear{2004})
\end{barticle}
\endbibitem

\bibitem{chimento1996exact}
\begin{barticle}
\bauthor{\bsnm{Chimento}, \binits{L.}},
\bauthor{\bsnm{Jakubi}, \binits{A.}}:
\batitle{Exact solutions and scalar fields in gravity}.
\bjtitle{Int. J. Mod. Phys. D}
\bvolume{5},
\bfpage{71}
(\byear{1996})
\end{barticle}
\endbibitem

\bibitem{kahya2015higher}
\begin{barticle}
\bauthor{\bsnm{Kahya}, \binits{E.}},
\bauthor{\bsnm{Khurshudyan}, \binits{M.}},
\bauthor{\bsnm{Pourhassan}, \binits{B.}},
\bauthor{\bsnm{Myrzakulov}, \binits{R.}},
\bauthor{\bsnm{Pasqua}, \binits{A.}}:
\batitle{Higher order corrections of the extended {C}haplygin gas cosmology
  with varying {G} and {$\Lambda$}}.
\bjtitle{The European Physical Journal C}
\bvolume{75}(\bissue{2}),
\bfpage{1}--\blpage{12}
(\byear{2015})
\end{barticle}
\endbibitem

\bibitem{jackiw2000particle}
\begin{botherref}
\oauthor{\bsnm{Jackiw}, \binits{R.}}:
A particle field theorist's lectures on supersymmetric, non-abelian fluid
  mechanics and d-branes.
arXiv preprint physics/0010042
(2000)
\end{botherref}
\endbibitem

\bibitem{pedram2008quantum}
\begin{barticle}
\bauthor{\bsnm{Pedram}, \binits{P.}},
\bauthor{\bsnm{Jalalzadeh}, \binits{S.}}:
\batitle{Quantum frw cosmological solutions in the presence of chaplygin gas
  and perfect fluid}.
\bjtitle{Physics Letters B}
\bvolume{659}(\bissue{1-2}),
\bfpage{6}--\blpage{13}
(\byear{2008})
\end{barticle}
\endbibitem

\bibitem{bento2002generalized}
\begin{barticle}
\bauthor{\bsnm{Bento}, \binits{M.}},
\bauthor{\bsnm{Bertolami}, \binits{O.}},
\bauthor{\bsnm{Sen}, \binits{A.A.}}:
\batitle{Generalized chaplygin gas, accelerated expansion, and
  dark-energy-matter unification}.
\bjtitle{Physical Review D}
\bvolume{66}(\bissue{4}),
\bfpage{043507}
(\byear{2002})
\end{barticle}
\endbibitem

\bibitem{gong2004constraints}
\begin{barticle}
\bauthor{\bsnm{Gong}, \binits{Y.}},
\bauthor{\bsnm{Duan}, \binits{C.-K.}}:
\batitle{Constraints on alternative models to dark energy}.
\bjtitle{Classical and Quantum Gravity}
\bvolume{21}(\bissue{15}),
\bfpage{3655}
(\byear{2004})
\end{barticle}
\endbibitem

\bibitem{carturan2003cosmological}
\begin{barticle}
\bauthor{\bsnm{Carturan}, \binits{D.}},
\bauthor{\bsnm{Finelli}, \binits{F.}}:
\batitle{Cosmological effects of a class of fluid dark energy models}.
\bjtitle{Physical Review D}
\bvolume{68}(\bissue{10}),
\bfpage{103501}
(\byear{2003})
\end{barticle}
\endbibitem

\bibitem{dev2004constraints}
\begin{barticle}
\bauthor{\bsnm{Dev}, \binits{A.}},
\bauthor{\bsnm{Jain}, \binits{D.}},
\bauthor{\bsnm{Alcaniz}, \binits{J.S.}}:
\batitle{Constraints on chaplygin quartessence from the class gravitational
  lens statistics and supernova data}.
\bjtitle{Astronomy \& Astrophysics}
\bvolume{417}(\bissue{3}),
\bfpage{847}--\blpage{852}
(\byear{2004})
\end{barticle}
\endbibitem

\bibitem{PhysRevD.69.083503}
\begin{barticle}
\bauthor{\bsnm{Bean}, \binits{R.}},
\bauthor{\bsnm{Dor\'e}, \binits{O.}}:
\batitle{Probing dark energy perturbations: The dark energy equation of state
  and speed of sound as measured by wmap}.
\bjtitle{Phys. Rev. D}
\bvolume{69},
\bfpage{083503}
(\byear{2004}).
\doiurl{10.1103/PhysRevD.69.083503}
\end{barticle}
\endbibitem

\bibitem{bilic2008transient}
\begin{botherref}
\oauthor{\bsnm{Bilic}, \binits{N.}},
\oauthor{\bsnm{Tupper}, \binits{G.B.}},
\oauthor{\bsnm{Viollier}, \binits{R.D.}}:
Transient acceleration in extended quartessence.
arXiv preprint astro-ph/0503428
(2008)
\end{botherref}
\endbibitem

\bibitem{fabris2002density}
\begin{barticle}
\bauthor{\bsnm{Fabris}, \binits{J.C.}},
\bauthor{\bsnm{Gon{\c{c}}alves}, \binits{S.V.}},
\bauthor{\bparticle{de} \bsnm{Souza}, \binits{P.E.}}:
\batitle{Density perturbations in a universe dominated by the chaplygin gas}.
\bjtitle{General Relativity and Gravitation}
\bvolume{34}(\bissue{1}),
\bfpage{53}--\blpage{63}
(\byear{2002})
\end{barticle}
\endbibitem

\bibitem{Yang_2019}
\begin{barticle}
\bauthor{\bsnm{Yang}, \binits{W.}},
\bauthor{\bsnm{Pan}, \binits{S.}},
\bauthor{\bsnm{Vagnozzi}, \binits{S.}},
\bauthor{\bsnm{Valentino}, \binits{E.D.}},
\bauthor{\bsnm{Mota}, \binits{D.F.}},
\bauthor{\bsnm{Capozziello}, \binits{S.}}:
\batitle{Dawn of the dark: unified dark sectors and the {EDGES} cosmic dawn
  21-cm signal}.
\bjtitle{Journal of Cosmology and Astroparticle Physics}
\bvolume{2019}(\bissue{11}),
\bfpage{044}--\blpage{044}
(\byear{2019}).
\doiurl{10.1088/1475-7516/2019/11/044}
\end{barticle}
\endbibitem

\bibitem{VCGPrime}
\begin{barticle}
\bauthor{\bsnm{Guo}, \binits{Z.-K.}},
\bauthor{\bsnm{Zhang}, \binits{Y.-Z.}}:
\batitle{Cosmology with a variable chaplygin gas}.
\bjtitle{Physics Letters B}
\bvolume{645}(\bissue{4}),
\bfpage{326}--\blpage{329}
(\byear{2007}).
\doiurl{10.1016/j.physletb.2006.12.063}
\end{barticle}
\endbibitem

\bibitem{Sethi_2006}
\begin{barticle}
\bauthor{\bsnm{Sethi}, \binits{G.}},
\bauthor{\bsnm{SINGH}, \binits{S.K.}},
\bauthor{\bsnm{KUMAR}, \binits{P.}},
\bauthor{\bsnm{JAIN}, \binits{D.}},
\bauthor{\bsnm{DEV}, \binits{A.}}:
\batitle{{VARIABLE} {CHAPLYGIN} {GAS}: {CONSTRAINTS} {FROM} {CMBR} {AND SNe}
  ia}.
\bjtitle{International Journal of Modern Physics D}
\bvolume{15}(\bissue{07}),
\bfpage{1089}--\blpage{1098}
(\byear{2006}).
\doiurl{10.1142/s0218271806008644}
\end{barticle}
\endbibitem

\bibitem{Zhai:2005mu}
\begin{barticle}
\bauthor{\bsnm{Zhai}, \binits{X.-H.}},
\bauthor{\bsnm{Xu}, \binits{Y.-D.}},
\bauthor{\bsnm{Li}, \binits{X.-Z.}}:
\batitle{{Viscous generalized Chaplygin gas}}.
\bjtitle{Int. J. Mod. Phys. D}
\bvolume{15},
\bfpage{1151}--\blpage{1162}
(\byear{2006})
{\href{https://arxiv.org/abs/astro-ph/0511814}{{arXiv:astro-ph/0511814}}}.
\doiurl{10.1142/S0218271806008784}
\end{barticle}
\endbibitem

\bibitem{PhysRevD.66.043507}
\begin{barticle}
\bauthor{\bsnm{Bento}, \binits{M.C.}},
\bauthor{\bsnm{Bertolami}, \binits{O.}},
\bauthor{\bsnm{Sen}, \binits{A.A.}}:
\batitle{Generalized chaplygin gas, accelerated expansion, and
  dark-energy-matter unification}.
\bjtitle{Phys. Rev. D}
\bvolume{66},
\bfpage{043507}
(\byear{2002}).
\doiurl{10.1103/PhysRevD.66.043507}
\end{barticle}
\endbibitem

\bibitem{sandvik2004end}
\begin{barticle}
\bauthor{\bsnm{Sandvik}, \binits{H.B.}},
\bauthor{\bsnm{Tegmark}, \binits{M.}},
\bauthor{\bsnm{Zaldarriaga}, \binits{M.}},
\bauthor{\bsnm{Waga}, \binits{I.}}:
\batitle{The end of unified dark matter?}
\bjtitle{Physical Review D}
\bvolume{69}(\bissue{12}),
\bfpage{123524}
(\byear{2004})
\end{barticle}
\endbibitem

\bibitem{del2013shear}
\begin{barticle}
\bauthor{\bsnm{Del~Popolo}, \binits{A.}},
\bauthor{\bsnm{Pace}, \binits{F.}},
\bauthor{\bsnm{Maydanyuk}, \binits{S.}},
\bauthor{\bsnm{Lima}, \binits{J.}},
\bauthor{\bsnm{Jesus}, \binits{J.}}:
\batitle{Shear and rotation in chaplygin cosmology}.
\bjtitle{Physical Review D}
\bvolume{87}(\bissue{4}),
\bfpage{043527}
(\byear{2013})
\end{barticle}
\endbibitem

\bibitem{sen2002tachyon}
\begin{barticle}
\bauthor{\bsnm{Sen}, \binits{A.}}:
\batitle{Tachyon matter}.
\bjtitle{Journal of High Energy Physics}
\bvolume{2002}(\bissue{07}),
\bfpage{065}
(\byear{2002})
\end{barticle}
\endbibitem

\bibitem{hannestad2002probing}
\begin{barticle}
\bauthor{\bsnm{Hannestad}, \binits{S.}},
\bauthor{\bsnm{M{\"o}rtsell}, \binits{E.}}:
\batitle{Probing the dark side: Constraints on the dark energy equation of
  state from cmb, large scale structure, and type ia supernovae}.
\bjtitle{Physical Review D}
\bvolume{66}(\bissue{6}),
\bfpage{063508}
(\byear{2002})
\end{barticle}
\endbibitem

\bibitem{guo2005parametrization}
\begin{barticle}
\bauthor{\bsnm{Guo}, \binits{Z.-K.}},
\bauthor{\bsnm{Ohta}, \binits{N.}},
\bauthor{\bsnm{Zhang}, \binits{Y.-Z.}}:
\batitle{Parametrization of quintessence and its potential}.
\bjtitle{Physical Review D}
\bvolume{72}(\bissue{2}),
\bfpage{023504}
(\byear{2005})
\end{barticle}
\endbibitem

\bibitem{caldwell2003phantom}
\begin{barticle}
\bauthor{\bsnm{Caldwell}, \binits{R.R.}},
\bauthor{\bsnm{Kamionkowski}, \binits{M.}},
\bauthor{\bsnm{Weinberg}, \binits{N.N.}}:
\batitle{Phantom energy: dark energy with w $<$ -1 causes a cosmic doomsday}.
\bjtitle{Physical review letters}
\bvolume{91}(\bissue{7}),
\bfpage{071301}
(\byear{2003})
\end{barticle}
\endbibitem

\bibitem{scolnic2018complete}
\begin{barticle}
\bauthor{\bsnm{Scolnic}, \binits{D.M.}},
\bauthor{\bsnm{Jones}, \binits{D.}},
\bauthor{\bsnm{Rest}, \binits{A.}},
\bauthor{\bsnm{Pan}, \binits{Y.}},
\bauthor{\bsnm{Chornock}, \binits{R.}},
\bauthor{\bsnm{Foley}, \binits{R.}},
\bauthor{\bsnm{Huber}, \binits{M.}},
\bauthor{\bsnm{Kessler}, \binits{R.}},
\bauthor{\bsnm{Narayan}, \binits{G.}},
\bauthor{\bsnm{Riess}, \binits{A.}}, \betal:
\batitle{The complete light-curve sample of spectroscopically confirmed sne ia
  from pan-starrs1 and cosmological constraints from the combined pantheon
  sample}.
\bjtitle{The Astrophysical Journal}
\bvolume{859}(\bissue{2}),
\bfpage{101}
(\byear{2018})
\end{barticle}
\endbibitem

\bibitem{suzuki2012hubble}
\begin{barticle}
\bauthor{\bsnm{Suzuki}, \binits{N.}},
\bauthor{\bsnm{Rubin}, \binits{D.}},
\bauthor{\bsnm{Lidman}, \binits{C.}},
\bauthor{\bsnm{Aldering}, \binits{G.}},
\bauthor{\bsnm{Amanullah}, \binits{R.}},
\bauthor{\bsnm{Barbary}, \binits{K.}},
\bauthor{\bsnm{Barrientos}, \binits{L.}},
\bauthor{\bsnm{Botyanszki}, \binits{J.}},
\bauthor{\bsnm{Brodwin}, \binits{M.}},
\bauthor{\bsnm{Connolly}, \binits{N.}}, \betal:
\batitle{The hubble space telescope cluster supernova survey. v. improving the
  dark-energy constraints above z $>$ 1 and building an early-type-hosted
  supernova sample}.
\bjtitle{The Astrophysical Journal}
\bvolume{746}(\bissue{1}),
\bfpage{85}
(\byear{2012})
\end{barticle}
\endbibitem

\bibitem{LIGODATA}
\begin{barticle}
\bauthor{\bsnm{{The LIGO Scientific Collaboration}}},
\bauthor{\bsnm{{the Virgo Collaboration}}},
\bauthor{\bsnm{{the KAGRA Collaboration}}}:
\batitle{Open data from the first and second observing runs of advanced ligo
  and advanced virgo}.
\bjtitle{SoftwareX}
\bvolume{13},
\bfpage{100658}
(\byear{2021}).
\doiurl{10.1016/j.softx.2021.100658}
\end{barticle}
\endbibitem

\bibitem{2021arXiv211103606T}
\begin{botherref}
\oauthor{\bsnm{{The LIGO Scientific Collaboration}}},
\oauthor{\bsnm{{the Virgo Collaboration}}},
\oauthor{\bsnm{{the KAGRA Collaboration}}}:
{GWTC-3: Compact Binary Coalescences Observed by LIGO and Virgo During the
  Second Part of the Third Observing Run}.
arXiv e-prints,
2111--03606
(2021)
{\href{https://arxiv.org/abs/2111.03606}{{arXiv:2111.03606}}}
{[gr-qc]}.
\doiurl{10.48550/arXiv.2111.03606}
\end{botherref}
\endbibitem

\bibitem{Torrado_2021}
\begin{barticle}
\bauthor{\bsnm{Torrado}, \binits{J.}},
\bauthor{\bsnm{Lewis}, \binits{A.}}:
\batitle{Cobaya: code for bayesian analysis of hierarchical physical models}.
\bjtitle{Journal of Cosmology and Astroparticle Physics}
\bvolume{2021}(\bissue{05}),
\bfpage{057}
(\byear{2021}).
\doiurl{10.1088/1475-7516/2021/05/057}
\end{barticle}
\endbibitem

\bibitem{2019ascl.soft10019T}
\begin{botherref}
\oauthor{\bsnm{{Torrado}}, \binits{J.}},
\oauthor{\bsnm{{Lewis}}, \binits{A.}}:
{Cobaya: Bayesian analysis in cosmology}.
Astrophysics Source Code Library, record ascl:1910.019
(2019)
\end{botherref}
\endbibitem

\bibitem{d2005fits}
\begin{botherref}
\oauthor{\bsnm{D'Agostini}, \binits{G.}}:
Fits, and especially linear fits, with errors on both axes, extra variance of
  the data points and other complications.
arXiv preprint physics/0511182
(2005)
\end{botherref}
\endbibitem

\bibitem{Paczynski1986}
\begin{barticle}
\bauthor{\bsnm{Paczynski}, \binits{B.}}:
\batitle{Gamma-ray bursts from neutron star mergers}.
\bjtitle{Astrophys. J.}
\bvolume{308},
\bfpage{43}--\blpage{46}
(\byear{1986})
\end{barticle}
\endbibitem

\bibitem{Planck2018}
\begin{botherref}
\oauthor{\bsnm{{Planck Collaboration}}}:
Planck 2018 results. vi. cosmological parameters.
Astron. Astrophys.
\textbf{641}
(2018)
{\href{https://arxiv.org/abs/1807.06209}{{1807.06209}}}.
\doiurl{10.1051/0004-6361/201833910}
\end{botherref}
\endbibitem

\bibitem{Kumar2015}
\begin{barticle}
\bauthor{\bsnm{Kumar}, \binits{P.}},
\bauthor{\bsnm{Zhang}, \binits{B.}}:
\batitle{The physics of gamma-ray bursts}.
\bjtitle{Phys. Rep.}
\bvolume{561},
\bfpage{1}--\blpage{109}
(\byear{2015})
\end{barticle}
\endbibitem

\bibitem{tanvir2009}
\begin{barticle}
\bauthor{\bsnm{Tanvir}, \binits{N.R.}},
\bauthor{\bsnm{Fox}, \binits{D.B.}},
\bauthor{\bsnm{Levan}, \binits{A.J.}},
\bauthor{\bsnm{Berger}, \binits{E.}},
\bauthor{\bsnm{Wiersema}, \binits{K.}},
\bauthor{\bsnm{Fynbo}, \binits{J.P.U.}},
\bauthor{\bsnm{Cucchiara}, \binits{A.}},
\bauthor{\bparticle{et} \bsnm{al.}}:
\batitle{A tidal disruption event as the origin of the gamma-ray burst grb
  090423}.
\bjtitle{Nature}
\bvolume{461},
\bfpage{1254}
(\byear{2009}).
\doiurl{10.1038/nature08459}
\end{barticle}
\endbibitem

\bibitem{cucchiara2011}
\begin{barticle}
\bauthor{\bsnm{Cucchiara}, \binits{A.}},
\bauthor{\bsnm{Levan}, \binits{A.J.}},
\bauthor{\bsnm{Strolger}, \binits{L.-G.}},
\bauthor{\bsnm{Berger}, \binits{E.}},
\bauthor{\bsnm{Fox}, \binits{D.B.}},
\bauthor{\bsnm{Tanvir}, \binits{N.R.}},
\bauthor{\bsnm{Brammer}, \binits{G.}},
\bauthor{\bparticle{et} \bsnm{al.}}:
\batitle{The afterglow of the first optically discovered gamma-ray burst
  afterglow: Grb 090423}.
\bjtitle{Astrophys. J.}
\bvolume{736},
\bfpage{1}
(\byear{2011})
\end{barticle}
\endbibitem

\bibitem{Rodney2015}
\begin{barticle}
\bauthor{\bsnm{Rodney}, \binits{S.A.}},
\bauthor{\bsnm{Strolger}, \binits{L.-G.}},
\bauthor{\bsnm{Tanvir}, \binits{N.R.}},
\bauthor{\bsnm{Berger}, \binits{E.}},
\bauthor{\bsnm{Cucchiara}, \binits{A.}},
\bauthor{\bsnm{Fox}, \binits{D.B.}},
\bauthor{\bsnm{Levan}, \binits{A.J.}},
\bauthor{\bparticle{et} \bsnm{al.}}:
\batitle{The optical afterglow of grb 140506a: constraints on the redshift and
  the nature of the progenitor}.
\bjtitle{Astrophys. J.}
\bvolume{801},
\bfpage{156}
(\byear{2015}).
\doiurl{10.1088/0004-637X/801/2/156}
\end{barticle}
\endbibitem

\bibitem{Narayan1992}
\begin{barticle}
\bauthor{\bsnm{Narayan}, \binits{R.}},
\bauthor{\bsnm{Paczynski}, \binits{B.}},
\bauthor{\bsnm{Piran}, \binits{T.}}:
\batitle{Gamma-ray bursts as the death throes of massive binary stars}.
\bjtitle{Astrophys. J.}
\bvolume{395},
\bfpage{83}
(\byear{1992})
\end{barticle}
\endbibitem

\bibitem{Woosley1993}
\begin{barticle}
\bauthor{\bsnm{Woosley}, \binits{S.E.}},
\bauthor{\bsnm{Langer}, \binits{N.}},
\bauthor{\bsnm{Weaver}, \binits{T.A.}}:
\batitle{The evolution and explosion of massive stars. i. numerical models of
  stellar evolution and nucleosynthesis from 8 to 100 solar masses}.
\bjtitle{Astrophys. J.}
\bvolume{405},
\bfpage{273}
(\byear{1993})
\end{barticle}
\endbibitem

\bibitem{MacFadyen1999}
\begin{barticle}
\bauthor{\bsnm{MacFadyen}, \binits{A.}},
\bauthor{\bsnm{Woosley}, \binits{S.E.}}:
\batitle{The physics of gamma-ray bursts}.
\bjtitle{Astrophys. J.}
\bvolume{524},
\bfpage{262}--\blpage{289}
(\byear{1999})
\end{barticle}
\endbibitem

\bibitem{Nagataki2007}
\begin{barticle}
\bauthor{\bsnm{Nagataki}, \binits{S.}},
\bauthor{\bsnm{Takahashi}, \binits{R.}},
\bauthor{\bsnm{Mizuta}, \binits{A.}},
\bauthor{\bsnm{Takiwaki}, \binits{T.}}:
\batitle{The prompt emission of gamma-ray bursts: implications of the grb
  060614 afterglow}.
\bjtitle{Astrophys. J.}
\bvolume{659},
\bfpage{512}--\blpage{529}
(\byear{2007})
\end{barticle}
\endbibitem

\bibitem{Nagataki2009}
\begin{barticle}
\bauthor{\bsnm{Nagataki}, \binits{S.}}:
\batitle{The afterglow of gamma-ray bursts: implications of the grb 060614
  afterglow}.
\bjtitle{Astrophys. J.}
\bvolume{704},
\bfpage{937}--\blpage{950}
(\byear{2009})
\end{barticle}
\endbibitem

\bibitem{Stanek2003}
\begin{barticle}
\bauthor{\bsnm{Stanek}, \binits{K.Z.}},
\bauthor{\bsnm{Matheson}, \binits{T.}},
\bauthor{\bsnm{Garnavich}, \binits{P.M.}},
\bauthor{\bsnm{Jha}, \binits{S.}},
\bauthor{\bsnm{Kirshner}, \binits{R.P.}},
\bauthor{\bsnm{Challis}, \binits{P.}},
\bauthor{\bsnm{Berlind}, \binits{P.}},
\bauthor{\bparticle{et} \bsnm{al.}}:
\batitle{The optical and near-infrared light curve of grb 030329: Evidence for
  a two-component afterglow}.
\bjtitle{Astrophys. J.}
\bvolume{591},
\bfpage{17}--\blpage{20}
(\byear{2003})
\end{barticle}
\endbibitem

\bibitem{MacFadyen2001}
\begin{barticle}
\bauthor{\bsnm{MacFadyen}, \binits{A.I.}},
\bauthor{\bsnm{Woosley}, \binits{S.E.}},
\bauthor{\bsnm{Heger}, \binits{A.}}:
\batitle{The physics of gamma-ray bursts}.
\bjtitle{Astrophys. J.}
\bvolume{550},
\bfpage{410}
(\byear{2001})
\end{barticle}
\endbibitem

\bibitem{Lattimer1976}
\begin{barticle}
\bauthor{\bsnm{Lattimer}, \binits{J.M.}},
\bauthor{\bsnm{Schramm}, \binits{D.N.}}:
\batitle{Neutron star-neutron star collisions and gamma-ray bursts}.
\bjtitle{Astrophys. J.}
\bvolume{210},
\bfpage{54}
(\byear{1976})
\end{barticle}
\endbibitem

\bibitem{Eichler1989}
\begin{barticle}
\bauthor{\bsnm{Eichler}, \binits{D.}},
\bauthor{\bsnm{Livio}, \binits{M.}},
\bauthor{\bsnm{Piran}, \binits{T.}},
\bauthor{\bsnm{Schramm}, \binits{D.N.}}:
\batitle{The cosmic fireball}.
\bjtitle{Nature}
\bvolume{340},
\bfpage{126}
(\byear{1989})
\end{barticle}
\endbibitem

\bibitem{Li1998}
\begin{barticle}
\bauthor{\bsnm{Li}, \binits{L.-X.}},
\bauthor{\bsnm{Paczyński}, \binits{B.}}:
\batitle{Transient events from gamma-ray bursts}.
\bjtitle{Astrophys. J.}
\bvolume{507},
\bfpage{59}--\blpage{62}
(\byear{1998})
\end{barticle}
\endbibitem

\bibitem{Rowlinson2014}
\begin{barticle}
\bauthor{\bsnm{Rowlinson}, \binits{A.}},
\bauthor{\bsnm{Gompertz}, \binits{B.P.}},
\bauthor{\bsnm{Dainotti}, \binits{M.G.}},
\bauthor{\bsnm{O'Brien}, \binits{P.T.}},
\bauthor{\bsnm{Wijers}, \binits{R.A.M.J.}},
\bauthor{\bparticle{van~der} \bsnm{Horst}, \binits{A.J.}}:
\batitle{The evolution of gamma-ray burst afterglows: a new empirical
  relationship}.
\bjtitle{Monthly Notices of the Royal Astronomical Society}
\bvolume{443},
\bfpage{1779}--\blpage{1787}
(\byear{2014})
\end{barticle}
\endbibitem

\bibitem{Rea2015}
\begin{barticle}
\bauthor{\bsnm{Rea}, \binits{N.}},
\bauthor{\bsnm{Gullón}, \binits{M.}},
\bauthor{\bsnm{Pons}, \binits{J.A.}},
\bauthor{\bsnm{Perna}, \binits{R.}},
\bauthor{\bsnm{Dainotti}, \binits{M.G.}},
\bauthor{\bsnm{Miralles}, \binits{J.A.}},
\bauthor{\bsnm{Torres}, \binits{D.F.}}:
\batitle{The long-term evolution of the gamma-ray burst afterglow grb 130427a:
  implications for the central engine}.
\bjtitle{Astrophys. J.}
\bvolume{813},
\bfpage{92}
(\byear{2015})
\end{barticle}
\endbibitem

\bibitem{Usov1992}
\begin{barticle}
\bauthor{\bsnm{{Usov}}, \binits{V.V.}}:
\batitle{{Magnetic monopoles and gamma-ray bursts}}.
\bjtitle{nat}
\bvolume{357},
\bfpage{472}--\blpage{474}
(\byear{1992}).
\doiurl{10.1038/357472a0}
\end{barticle}
\endbibitem

\bibitem{Liang2018}
\begin{barticle}
\bauthor{\bsnm{{Liang}}, \binits{L.}}:
\batitle{{On the Origin of the Prompt Emission in Gamma-Ray Bursts}}.
\bjtitle{apjs}
\bvolume{236}(\bissue{2}),
\bfpage{26}
(\byear{2018}).
\doiurl{10.3847/1538-4365/aad642}
\end{barticle}
\endbibitem

\bibitem{Ai2018}
\begin{barticle}
\bauthor{\bsnm{{Ai}}, \binits{S.}}:
\batitle{{The Photospheric Emission Model for Gamma-Ray Bursts: A Comprehensive
  Study}}.
\bjtitle{apj}
\bvolume{860}(\bissue{1}),
\bfpage{57}
(\byear{2018}).
\doiurl{10.3847/1538-4357/aac90c}
\end{barticle}
\endbibitem

\bibitem{Dainotti2008}
\begin{barticle}
\bauthor{\bsnm{{Dainotti}}, \binits{M.G.}}:
\batitle{{A new distance indicator for gamma-ray bursts}}.
\bjtitle{mnras}
\bvolume{391}(\bissue{1}),
\bfpage{79}--\blpage{83}
(\byear{2008}).
\doiurl{10.1111/j.1745-3933.2008.00560.x}
\end{barticle}
\endbibitem

\bibitem{Dainotti2016}
\begin{barticle}
\bauthor{\bsnm{{Dainotti}}, \binits{M.G.}}:
\batitle{{A Fundamental Plane for Long Gamma-Ray Bursts with X-ray Plateaus}}.
\bjtitle{apjl}
\bvolume{825}(\bissue{2}),
\bfpage{6}
(\byear{2016}).
\doiurl{10.3847/2041-8205/825/2/L20}
\end{barticle}
\endbibitem

\bibitem{Dainotti2017c}
\begin{barticle}
\bauthor{\bsnm{{Dainotti}}, \binits{M.G.}}:
\batitle{{A New Fundamental Plane for Long Gamma-Ray Bursts with X-ray Plateaus
  and its Cosmological Implications}}.
\bjtitle{apj}
\bvolume{848}(\bissue{2}),
\bfpage{88}
(\byear{2017}).
\doiurl{10.3847/1538-4357/aa8856}
\end{barticle}
\endbibitem

\bibitem{Dago05}
\begin{barticle}
\bauthor{\bsnm{{D'Agostini}}, \binits{G.}}:
\batitle{{Fits, and especially linear fits, with errors on both axes, extra
  variance of the data points and other complications}}.
\bjtitle{ArXiv e-prints}
(\byear{2005})
{\href{https://arxiv.org/abs/physics/0511182}{{physics/0511182}}}.
\doiurl{10.48550/arXiv.physics/0511182}
\end{barticle}
\endbibitem

\bibitem{dainotti2023gamma}
\begin{barticle}
\bauthor{\bsnm{Dainotti}, \binits{M.}},
\bauthor{\bsnm{Lenart}, \binits{A.{\L}.}},
\bauthor{\bsnm{Chraya}, \binits{A.}},
\bauthor{\bsnm{Sarracino}, \binits{G.}},
\bauthor{\bsnm{Nagataki}, \binits{S.}},
\bauthor{\bsnm{Fraija}, \binits{N.}},
\bauthor{\bsnm{Capozziello}, \binits{S.}},
\bauthor{\bsnm{Bogdan}, \binits{M.}}:
\batitle{The gamma-ray bursts fundamental plane correlation as a cosmological
  tool}.
\bjtitle{Monthly Notices of the Royal Astronomical Society}
\bvolume{518}(\bissue{2}),
\bfpage{2201}--\blpage{2240}
(\byear{2023})
\end{barticle}
\endbibitem

\bibitem{HER-AL}
\begin{barticle}
\bauthor{\bsnm{"Hern\'andez-Almada}, \binits{A.}},
\bauthor{\bsnm{Garc\'\i{}a-Aspeitia}, \binits{M.A.}},
\bauthor{\bsnm{Rodr\'\i{}guez-Meza}, \binits{M.A.}},
\bauthor{\bsnm{Motta}, \binits{V..}}:
\batitle{{A hybrid model of viscous and Chaplygin gas to tackle the Universe
  acceleration}}.
\bjtitle{Eur. Phys. J. C}
\bvolume{81}(\bissue{4}),
\bfpage{295}
(\byear{2021})
{\href{https://arxiv.org/abs/2103.16733}{{arXiv:2103.16733}}}
{[astro-ph.CO]}.
\doiurl{10.1140/epjc/s10052-021-09104-w}
\end{barticle}
\endbibitem

\end{thebibliography}

\end{document}